\documentclass[submission,copyright,creativecommons]{eptcs} 
\providecommand{\event}{ML 2016}
\usepackage[utf8]{inputenc}

\title{Sundials/ML: Connecting OCaml to the Sundials Numeric Solvers}
\author{
  Timothy Bourke
  \institute{Inria Paris}
  \institute{École normale supérieure, PSL University}
  \email{tim@tbrk.org}
\and
  Jun Inoue
  \institute{National Institute of Advanced Industrial Science and Technology}
  \email{jun.inoue@aist.go.jp}
\and
  Marc Pouzet
  \institute{Sorbonne Universités, UPMC Univ Paris 06}
  \institute{École normale supérieure, PSL University}
  \institute{Inria Paris}
  \email{marc.pouzet@ens.fr}
} 
\def\titlerunning{Sundials/ML}
\def\authorrunning{T. Bourke, J. Inoue \& M. Pouzet}

\hypersetup{%
  pdftitle={Sundials/ML: Connecting OCaml to the Sundials Numeric 
  Solvers},
  pdfauthor={Timothy Bourke \& Jun Inoue \& Marc Pouzet},
  pdfsubject={Interfacing OCaml to the Sundials C library},
  pdfkeywords={Foreign Function Interfaces; OCaml; Numeric solvers;
               Applications of static typing},
}

\usepackage[T1]{fontenc}
\usepackage{textcomp}

\usepackage[binary-units,detect-mode]{siunitx}

\usepackage{amsmath}
\usepackage{physics}

\usepackage[final]{graphicx}

\usepackage{tikz}
\usetikzlibrary{arrows,automata,positioning,calc,plotmarks,
  decorations.pathreplacing}

\usepackage{bbding}

\usepackage{url}
\usepackage[printonlyused]{acronym}
\usepackage{fancyvrb}
\usepackage{upquote}

\DefineVerbatimEnvironment{Verbatim}{Verbatim}
    {commandchars=\\\{\},xleftmargin=1.2em,fontsize=\small}
\DefineVerbatimEnvironment{NumberedVerbatim}{Verbatim}
    {commandchars=\\\{\},xleftmargin=1.2em,numbers=left,
     numbersep=5pt,numberblanklines=false,
     fontsize=\small}
\newcommand{\mytilde}{{\raisebox{-.7ex}{\textasciitilde}}} 
\RecustomVerbatimCommand{\verb}{Verb}{commandchars=\\\{\}}

\usepackage[nameinlink,noabbrev]{cleveref} 
\makeatletter 
\newcommand{\Creflabel}[1]{\csname Cref@#1@name\endcsname}
\newcommand{\Creflabels}[1]{\csname Cref@#1@name@plural\endcsname}
\newcommand{\creflabel}[1]{\csname cref@#1@name\endcsname}
\newcommand{\creflabels}[1]{\csname cref@#1@name@plural\endcsname}
\makeatother

\usepackage[inline]{enumitem}
\newlist{inparaenum}{enumerate*}{1}
\setlist[inparaenum,1]{label=({\roman*}),ref=\roman*}
\newlist{stepenum}{enumerate*}{1}
\setlist[stepenum,1]{label=({\it\roman*}),ref=\emph{\roman*}}
\crefname{stepenumi}{step}{steps}

\newcommand{\cvode}{\textsc{\it CVODE}}
\newcommand{\cvodes}{\textsc{\it CVODES}}
\newcommand{\ida}{\textsc{\it IDA}}
\newcommand{\idas}{\textsc{\it IDAS}}
\newcommand{\kinsol}{\textsc{\it KINSOL}}
\newcommand{\arkode}{\textsc{\it ARKODE}}

\newcommand{\filename}[1]{\textsf{#1}}

\acrodef{API}{Application Programming Interface}
\acrodef{BBD}{Band-Block-Diagonal}
\acrodef{CSC}{Compressed-Sparse-Column}
\acrodef{CSR}{Compressed-Sparse-Row}
\acrodef{DAE}{differential algebraic equation}
\acrodef{DLS}{Direct Linear Solver}
\acrodefplural{DLS}[DLS]{Direct Linear Solvers}
\acrodef{FFI}{Foreign Function Interface}
\acrodef{GADT}{Generalized Abstract Data Type}
\acrodef{MPI}{Message Passing Interface}
\acrodef{ODE}{ordinary differential equation}
\acrodef{PCG}{Preconditioned Conjugate Gradient}
\acrodef{SPBCGS}{Scaled Preconditioned Biconjugate Stabilized}
\acrodef{SPILS}{Scaled Preconditioned Iterative Linear Solver}
\acrodefplural{SPILS}[SPILS]{Scaled Preconditioned Iterative Linear Solvers}
\acrodef{SPFGMR}{Scaled Preconditioned Flexible Generalized Minimum 
Residual}
\acrodef{SPGMR}{Scaled Preconditioned Generalized Minimum Residual}
\acrodef{SPTFQMR}{Scaled Preconditioned Transpose-Free Quasi-Minimal 
Residual}
\acrodef{SLS}{Sparse Linear Solver}
\acrodefplural{SLS}[SLS]{Sparse Linear Solvers}

\begin{document}

\maketitle
\begin{abstract} 
This paper describes the design and implementation of a comprehensive OCaml 
interface to the Sundials library of numeric solvers for ordinary 
differential equations, differential algebraic equations, and non-linear 
equations.
The interface provides a convenient and memory-safe alternative to using 
Sundials directly from~C and facilitates application development by 
integrating with higher-level language features, like garbage-collected 
memory management, algebraic data types, and exceptions.
Our benchmark results suggest that the interface overhead is acceptable: the 
standard examples are rarely twice as slow in OCaml than in C, and often 
less than 50\% slower.
The challenges in interfacing with Sundials are to efficiently and safely 
share data structures between OCaml and C, to support multiple 
implementations of vector operations, matrices, and linear solvers through a 
common interface, and to manage calls and error signalling to and from 
OCaml.
We explain how we overcame these difficulties using a combination of 
standard techniques such as phantom types and polymorphic variants, and 
carefully crafted data representations.
%
\end{abstract} 
\section{Introduction}\label{sec:intro}

Sundials~\cite{HindmarshEtAl:Sundials:2005} is a suite of six numeric 
solvers:
\cvode{}, for \aclp{ODE},
\cvodes{}, adding support for quadrature integration and sensitivity 
analysis,
\ida{}, for \aclp{DAE},
\idas{}, adding support for quadrature integration and sensitivity analysis,
\arkode{}, for \aclp{ODE} using adaptive-step additive Runge-Kutta methods, 
and \kinsol{}, for non-linear equations.
The six solvers share data structures and operations for vectors, matrices, 
and linear solver routines.
They are implemented in the C language.

In this article we describe the design and implementation of a comprehensive 
OCaml~\cite{LeroyEtAl:OCamlMan:2018} interface to the Sundials library, 
which we call \emph{Sundials/ML}.
The authors of Sundials, Hindmarsh \emph{et 
al.}~\cite{Sundials:Cvode:3.1.0}, describe ``a general movement away from 
Fortran and toward C in scientific computing'' and note both the utility of 
C's pointers, structures, and dynamic memory management for writing such 
software, and also the availability, efficiency, and relative ease of 
interfacing to it from Fortran.
So, why bother writing an interface from OCaml?
We think that OCaml interfaces to libraries like Sundials are ideal for
\begin{inparaenum}
\item
programs that mix numeric computation with symbolic manipulation, like 
interpreters and simulators for hybrid modelling languages;
\item
rapidly developing complex numerical models; and
\item
incorporating numerical approximation into general-purpose applications.
\end{inparaenum}
Compared to C, OCaml detects many mistakes through a combination of static 
analyses (strong typing) and dynamic checks (for instance, array bounds 
checking), manages memory automatically using garbage collection, and 
propagates errors as exceptions rather than return codes.
Not only does the OCaml type and module system detect a large class of 
programming errors, it also enables rich library interfaces that clarify and 
enforce correct use.
We exploit this possibility in our library; for example, algebraic data 
types are used to structure solver configuration as opposed to multiple 
interdependent function calls, polymorphic typing ensures consistent 
creation and use of sessions and linear solvers, and phantom types are 
applied to enforce restrictions on vector and matrix use.
On the other hand, all such interfaces add additional code and thus runtime 
overhead and the possibility of bugs---we discuss these issues in 
\cref{sec:eval}.

The basic techniques for interfacing OCaml and C are well 
understood~\cite[Chapter~20]{LeroyEtAl:OCamlMan:2018}%
\cite{Monnier:OcamlandC:2013}%
\cite[Chapter~12]{ChaillouxManPag:ObjCaml:2000b}%
\cite[Chapters~19--21]{MinskyMadHic:RWOCaml:2013}, and it only takes one or 
two weeks to make a working interface to one of the solvers using the basic 
array-based vector data structure.
But not all interfaces are created equal!

It takes much longer to treat the full range of features available in the 
Sundials suite, like the two solvers with quadrature integration and 
sensitivity analysis, the different linear solver modules and their 
associated matrix representations, and the diverse vector data structures.
In particular, it was not initially clear how to provide all of these 
features in an integrated way with minimal code duplication and good support 
from the OCaml type system.
\Cref{sec:tech} presents our solution to this problem.
The intricacies of the Sundials library called for an especially careful 
design, particularly in the memory layout of a central vector data structure 
which took several iterations to perfect.
The key challenges are to limit copying by modifying data in place, to 
interact correctly with the garbage collector to avoid memory leaks and data 
corruption, and to exploit the type and module systems to express documented 
constraints and provide a convenient interface.
Our interface employs higher-order functions and currying, but such 
functional programming techniques are incidental; ultimately, we cannot 
escape the imperative nature of the underlying library.
Similarly, in several instances static types are unable to encode 
constraints that change with the solver's state, and we are obliged to add 
dynamic checks to provide a safe and truly high-level interface.

Otherwise, a significant engineering effort is required to create a robust 
build system able to support the various configurations and versions of 
Sundials on different platforms, and to translate and debug the 100-odd 
standard examples that were indispensable for designing, evaluating, and 
testing our interface.
\Cref{sec:eval} summarizes the performance results produced through this 
effort.

The Sundials/ML interface is used in the runtime of the Zélus programming 
language~\cite{BourkePou:HSCC:2013}.
Zélus is a synchronous language~\cite{BenvenisteEtAl:12Years:2003} extended 
with \acp{ODE} for programming embedded systems and modelling their 
environments.
Its compiler generates OCaml code to be executed together with a simulation 
runtime that orchestrates discrete computations and the numerical 
approximation of continuous trajectories.

The remainder of the paper is structured as follows. First, we describe the 
overall design of Sundials and Sundials/ML from a library user's 
point-of-view with example programs (\cref{sec:overview}).
Then we explain the main technical challenges that we overcame and the 
central design choices (\cref{sec:tech}).
Finally, we present an evaluation of the performance of our binding 
(\cref{sec:eval}) before concluding (\cref{sec:concl}).

The documentation and source code---approximately \num{15000} lines of 
OCaml, and \num{17000} lines of C, not counting a significant body of 
examples and tests---of the interface described in this paper is available 
under a 3-clause BSD licence at 
\url{http://inria-parkas.github.io/sundialsml/}.
The code has been developed intermittently over eight years and adapted for 
successive Sundials releases through to the current 3.1.0 version.

\section{Overview}\label{sec:overview}

In this section, we outline the \acp{API} of Sundials and Sundials/ML and
describe their key data structures.
We limit ourselves to the elements necessary to explain and justify 
subsequent technical points.
A complete example is presented at the end of the section.

\subsection{Overview of Sundials}\label{sec:coverview} 


A mathematical description of Sundials can be found in Hindmarsh \emph{et 
al.}~\cite{HindmarshEtAl:Sundials:2005}.
The user manuals\footnote{https://computation.llnl.gov/casc/sundials/}
give a thorough overview of the library and the details of every function.
The purposes of the four basic solvers are readily summarized:
\begin{itemize}
\item
  \cvode{}  approximates $Y(t)$ from $\dot{Y} = f(Y)$ and $Y(t_0) = Y_0$.
\item
  \ida{}    approximates $Y(t)$ from $F(Y, \dot{Y}) = 0$, $Y(t_0) = Y_0$,
            and $\dot{Y}(t_0) = \dot{Y}_0$.
\item
  \arkode{} approximates $Y(t)$
            from $M\dot{Y} = f_E(Y) + f_I(Y)$ and $Y(t_0) = Y_0$.
\item
  \kinsol{} calculates $U$ from $F(U) = 0$ and initial guess $U_0$.
\end{itemize}
The problems are stated over vectors $Y\!$, $\dot{Y}\!$, and $U\!$, and 
expressed in terms of a function~$f$ from states to derivatives, a 
function~$F$ from states and derivatives to a residual, or the combination 
of an explicit function~$f_E$ and an implicit function~$f_I$, both from 
states to derivatives, together with a mapping matrix~$M$.
The first three solvers find solutions that depend on an independent 
variable~$t$, usually considered to be the simulation time.
The two solvers that are not mentioned above, \cvodes{} and \idas{}, 
essentially introduce a set of parameters $P$---giving models $f(Y, P)$ and 
$F(Y, \dot{Y}, P)$---and permit the calculation of parameter sensitivities, 
$S(t) = \frac{\partial Y(t)}{\partial P}$ using a variety of different 
techniques.

Four features are most relevant to the OCaml interface: solver sessions, 
vectors, matrices, and linear solvers.

\subsubsection{Solver sessions}\label{sec:coverview:sessions} 

\begin{figure}
\begin{center} 
\begin{NumberedVerbatim}
cvode_mem = CVodeCreate(CV_BDF, CV_NEWTON);\label{cvodeinit:create}
if(check_flag((void *)cvode_mem, "CVodeCreate", 0)) return(1);

flag = CVodeSetUserData(cvode_mem, data);\label{cvodeinit:userdata}
if(check_flag(&flag, "CVodeSetUserData", 1)) return(1);

flag = CVodeInit(cvode_mem, f, T0, u);\label{cvodeinit:init}
if(check_flag(&flag, "CVodeInit", 1)) return(1);

flag = CVodeSStolerances(cvode_mem, reltol, abstol);\label{cvodeinit:tol}
if (check_flag(&flag, "CVodeSStolerances", 1)) return(1);

LS = SUNSPGMR(u, PREC_LEFT, 0);\label{cvodeinit:spgmr}
if(check_flag((void *)LS, "SUNSPGMR", 0)) return(1);

flag = CVSpilsSetLinearSolver(cvode_mem, LS);\label{cvodeinit:setlinear}
if (check_flag(&flag, "CVSpilsSetLinearSolver", 1)) return 1;

flag = CVSpilsSetJacTimes(cvode_mem, jtv);\label{cvodeinit:jactimes}
if(check_flag(&flag, "CVSpilsSetJacTimes", 1)) return(1);

flag = CVSpilsSetPreconditioner(cvode_mem, Precond, PSolve);\label{cvodeinit:precond}
if(check_flag(&flag, "CVSpilsSetPreconditioner", 1)) return(1);
\end{NumberedVerbatim}
\end{center} 
\caption{Extract from the cvDiurnal\_kry example in C, distributed with 
Sundials~\cite{HindmarshEtAl:Sundials:2005}.}\label{fig:cvodeinit:c}
\end{figure}

Using any of the Sundials solvers involves the same basic pattern:
\begin{stepenum}
\item\label{step:create}
a session object is created;
\item\label{step:init}
several inter-dependent functions are called to initialize the session;
\item\label{step:set}
``set\,$\ast$'' functions are called to give parameter values;
\item\label{step:solve}
a ``solve'' or ``step'' function is called repeatedly to approximate a 
solution;
\item\label{step:get}
``get$\,\ast$'' functions are called to retrieve results;
\item
the session and related data structures are freed.
\end{stepenum}
The sequence in~\cref{fig:cvodeinit:c}, extracted from an example 
distributed with Sundials, is typical of 
\cref*{step:create,step:init,step:set}.
Line~\ref{cvodeinit:create}
creates a session with the (\cvode{}) solver, that is, an abstract type 
implemented as a pointer to a structure containing solver parameters and 
state that is passed to and manipulated by all subsequent calls.
This function call, and all the others, are followed by statements that 
check return codes.
Line~\ref{cvodeinit:userdata} specifies a ``user data'' pointer that is 
passed by the solver to all callback functions to provide session-local 
storage.
Line~\ref{cvodeinit:init} specifies a callback function~\verb"f" that 
defines the problem to solve, here a function from a vector of variable 
values to their derivatives ($\dot{X} = f(X)$), an initial value~\verb"T0" 
for the independent variable, and a vector of initial variable 
values~\verb"u" that implicitly defines the problem size (\mbox{$X_0 = u$}).
The other calls specify tolerance values (line~\ref{cvodeinit:tol}), 
instantiate an iterative linear solver (line~\ref{cvodeinit:spgmr}), attach 
it to the solver session (line~\ref{cvodeinit:setlinear}), and set callback 
functions \verb"jtv", defining the problem Jacobian 
(line~\ref{cvodeinit:jactimes}), and \verb"Precond" and \verb"PSolve", 
defining preconditioning
(line~\ref{cvodeinit:precond}).
The loop that calculates values of $X$ over time, and the functions that 
retrieve solver results and statistics, and free memory are not shown.

While the \ida{} and \kinsol{} solvers follow the same pattern, using the 
\cvodes{} and \idas{} solvers is a bit more involved.
These solvers provide additional calls that augment a basic solver session 
with features for more efficiently calculating certain integrals and for 
analyzing the sensitivity of a solution to changes in model parameters using 
either so called forward methods or adjoint methods.
The adjoint methods involve solving an \ac{ODE} or \ac{DAE} problem by first 
integrating normally, and then initializing new ``backward'' sessions that 
are integrated in the reverse direction.

The routines that initialize and configure solver sessions are subject to 
rules constraining their presence, order, and parameters.
For instance, in the example, the call to \verb"CVodeSStolerances" must 
follow the call to \verb"CVodeInit" and precede calls to the step function; 
calling \verb"CVodeCreate" with the parameter \verb"CV_NEWTON" necessitates 
calls to configure a linear solver; and the \verb"CVSpilsSetLinearSolver" 
call requires an iterative linear solver such as \verb"SUNSPGMR" which, in 
turn, requires a call to \verb"CVSpilsSetJacTimes" and, since the 
\verb"PREC\_LEFT" argument is given, a call to 
\verb"CVSpilsSetPreconditioner" with at least a \verb"PSolve" value.

\subsubsection{Vectors}\label{sec:coverview:vectors} 

The manipulation of vectors is fundamental in Sundials.
Vectors of floating-point values are represented as an abstract data type 
termed an \emph{nvector} which combines a data pointer and 26 function 
pointers to operations that manipulate the data.
Typical of these operations are \emph{nvlinearsum}, which calculates the 
scaled sum of nvectors~$X$ and~$Y$ into a third vector~$Z$, that is, $Z = aX 
+ bY\!$, and \emph{nvmaxnorm}, which returns the maximum absolute value of 
an nvector~$X$.
Nvectors must also provide an \emph{nvclone} operation that produces a new 
nvector of the same size and with the same operations as an existing one.
Solver sessions are seeded with an initial nvector that they clone 
internally and manipulate solely through the abstract operations---they are 
thus defined in a data-independent manner.

Sundials provides eight instantiations of the nvector type: serial
nvectors, parallel nvectors, OpenMP nvectors, Pthreads nvectors, Hypre 
ParVector nvectors, PETSC nvectors, RAJA nvectors, and CUDA nvectors.
Serial nvectors store and manipulate arrays of floating-point values.
Parallel nvectors store a local array of floating-point values and a
\ac{MPI} communicator; some operations, like \emph{nvlinearsum},
simply loop over the local array, while others, like \emph{nvmaxnorm},
calculate locally and then synchronize via a global reduce operation.
OpenMP nvectors and Pthreads nvectors operate concurrently on arrays
of floating-point values.
Despite the lack of real multi-threading support in the OCaml runtime, 
binding to these nvector routines is unproblematic since they do not call 
back into user code.
The serial, OpenMP, and Pthreads nvectors provide operations for accessing 
the underlying data array directly; this feature is notably exploited in the 
implementation of certain linear solvers.
The Hypre ParVector, PETSC, and RAJA nvectors interface with scientific 
computing libraries that have not been ported to OCaml; they are not 
supported by Sundials/ML.
We have not yet attempted to support CUDA nvectors.
Finally, library users may also provide their own \emph{custom nvectors} by 
implementing the set of basic operations over a custom representation.

\subsubsection{Matrices}\label{sec:coverview:matrices} 

In recent versions of Sundials, matrices are treated like nvectors: they are 
implemented as an abstract data type combining a data pointer with pointers 
to 9 abstract operations~\cite[\textsection{}7]{Sundials:Cvode:3.1.0}.
There are, for instance, operations to clone, to destroy, to scale and add 
to another matrix, and to multiply by an nvector.

Implementations are provided for two-dimensional dense, banded, and sparse 
matrices, and it is possible for users to define custom matrices by 
implementing the abstract operations.
The \emph{matrix content} of dense matrices consists of fields for the 
matrix dimensions and data length, a data array of floating-point values, 
and an array of pre-calculated column pointers into the data array.
The content of banded matrices, which only contain the main diagonal and a 
certain number of diagonals above and below it, are represented similarly 
but with extra fields to record the numbers of diagonals.
The content of sparse matrices includes the number of non-zero elements that 
can potentially be stored, a data array of non-zero elements, and two 
additional integer arrays.
The interpretation of the integer arrays depends on a storage format field.
For \ac{CSC} matrices, an \emph{indexptrs} array maps column numbers (from 
zero) to indices of the data array and an \emph{indexvals} array maps 
indices of the data array to row numbers (from zero).
So, for instance, the non-zero elements of the $(j-1)$th~column are stored 
consecutively in the data array at indices~$\mathit{indexptrs}[j] \le k < 
\mathit{indexptrs}[j+1]$, and the row of each element 
is~$\mathit{indexvals}[k] + 1$.
For \ac{CSR} matrices, \emph{indexptrs} maps row numbers to data indices and 
\emph{indexvals} maps data indices to column numbers.

In the interests of calculation speed, the different Sundials routines often 
violate the abstract interface and access the underlying representations 
directly.
This induces compatibility constraints: for instance, dense matrices may 
only be multiplied with serial, OpenMP, and Pthreads nvectors, and the KLU 
linear solver only manipulates sparse matrices.
In \cref{sec:matrices}, we explain how these rules are expressed as typing 
constraints in the OCaml interface.

\subsubsection{Linear solvers}\label{sec:coverview:lsolvers} 

Each of the solvers must resolve non-linear algebraic systems.
For this, they use either ``functional iteration'' (\cvode{} and \cvodes{} 
only), or, usually, `Newton iteration'.
Newton iteration, in turn, requires the solution of linear systems.

Several linear solvers are provided with Sundials.
They are instantiated by generic routines, like \verb"SUNSPGMR" in the 
example, and attached to solver sessions.
Generic linear solvers factor out common algorithms that solver-specific 
linear solvers specialize and combine with callback routines, like 
\verb"jtv", \verb"Precond", and \verb"Psolve" in the example.
Internally, generic linear solvers are implemented, similarly to nvectors 
and matrices, as an abstract data type combining a data pointer with 13 
function pointers.
Solver-specific linear solvers are invoked through a generic interface 
comprising four function pointers in a session object: \emph{linit}, 
\emph{lsetup}, \emph{lsolve}, and \emph{lfree}.
Users may provide their own \emph{custom} linear solvers by specifying 
subsets of the 13 operations, or even their own \emph{alternate} linear 
solvers by directly setting the four pointers.

Sundials includes three main linear solver families: a diagonal 
approximation of system Jacobians using difference quotients, \acp{DLS} that 
perform LU factorization on Jacobian matrices, and
\acp{SPILS} based on Krylov methods.
The \ac{DLS} modules require callback routines to calculate an explicit 
representation of the Jacobian matrix of a system.
The \ac{SPILS} modules require callback routines to multiply vectors by an 
(implicit) Jacobian matrix and to precondition.

As was the case for solver sessions, the initialization and use of linear 
solvers is subject to various rules.
For instance, the \ac{DLS} modules exploit the underlying representation of 
serial, OpenMP, and Pthreads nvectors, and they cannot be used with other 
nvectors.
The \ac{SPILS} modules combine a method (\acs{SPGMR}, \acs{SPFGMR}, 
\acs{SPBCGS}, \acs{SPTFQMR}, or \acs{PCG}) with an optional preconditioner.
There are standard preconditioners (left, right, or both) for which users 
supply a solve function and, optionally, a setup function, and which work 
for any nvector, and also a banded matrix preconditioner that is only 
compatible with serial nvectors, and a \ac{BBD} preconditioner that is only 
compatible with parallel nvectors.
The encoding of these restrictions into the OCaml interface is described in 
\cref{sec:linsolv}.

\subsection{Overview of Sundials/ML}\label{sec:mloverview} 

The structure of the OCaml interface mostly follows that of the underlying 
library and values are consistently and systematically renamed: 
module-identifying prefixes are replaced by module paths and words beginning 
with upper-case letters are separated by underscores and put into 
lower-case.
For instance, the function name \verb"CVSpilsSetGSType" becomes
\verb"Cvode.Spils.set_gs_type".
This makes it easy to read the original documentation and to adapt existing 
source code, like, for instance, the examples provided with Sundials.
We did, however, make several changes both for programming convenience and 
to increase safety, namely:
\begin{inparaenum}
\item
solver sessions are mostly configured via algebraic data types rather than 
multiple function calls;
\item
errors are signalled by exceptions rather than return codes;
\item
user data is shared between callback routines via partial function 
applications (closures);
\item
vectors are checked for compatibility using a combination of static and 
dynamic checks; and
\item
explicit free commands are not necessary since OCaml is a garbage-collected 
language.
\end{inparaenum}

\SaveVerb{precond}"precond"
\SaveVerb{jtv}"jtv"
\SaveVerb{psolve}"psolve"
\SaveVerb{f}"f"

\SaveVerb{data}"data"
\SaveVerb{tzero}"t0"
\SaveVerb{u}"u"
\SaveVerb{reltol}"reltol"
\SaveVerb{abstol}"abstol"
\begin{figure}
\begin{center} 
\begin{NumberedVerbatim}
let cvode_mem =
  Cvode.(init BDF
    (Newton Spils.(solver (spgmr u)
                          \mytilde{}jac_times_vec:(None, jtv data)
                          (prec_left \mytilde{}setup:(precond data) (psolve data))))
    (SStolerances (reltol, abstol))
    (f data) t0 u)
in
\end{NumberedVerbatim}
\end{center} 
\caption{Extract from our OCaml adaptation of cvDiurnal\_kry.
The definitions of the \protect\UseVerb{precond}, \protect\UseVerb{jtv}, 
\protect\UseVerb{psolve}, and \protect\UseVerb{f} functions and those of the 
\protect\UseVerb{data}, \protect\UseVerb{reltol}, \protect\UseVerb{abstol}, 
\protect\UseVerb{tzero}, and \protect\UseVerb{u} variables are not 
shown.}\label{fig:cvodeinit:ocaml}
\end{figure}

The OCaml program extract in \cref{fig:cvodeinit:ocaml} is functionally 
equivalent to the C code of \cref{fig:cvodeinit:c}.
%
But rather than specifying Newton iteration by passing a constant 
(\verb"CV_NEWTON") and later calling linear solver routines (like 
\verb"CVSpilsSetLinearSolver"), a solver session is configured by passing a 
value that contains all the necessary parameters.
This makes it impossible to specify Newton iteration without also properly 
configuring a linear solver.
The given value is translated by the interface into the correct sequence of 
calls to Sundials.
The interface checks the return code of each function call and raises an 
exception if necessary.
In the extract, the \verb"\mytilde{}setup" and 
\verb"\mytilde{}jac_times_vec" markers denote labelled 
arguments~\cite[\textsection 
4.1]{LeroyEtAl:OCamlMan:2018}; we use them for optional arguments, as in 
this example, and also to clarify the meaning of multiple arguments of the 
same type.
The callback functions---\UseVerb{precond}, \UseVerb{jtv}, \UseVerb{psolve}, 
and \UseVerb{f}---are all applied to \UseVerb{data}.
This use of partial application over a shared value replaces the ``user 
data'' mechanism of Sundials (\verb"CVodeSetUserData") that provides
session-local storage; it is more natural in OCaml and it frees the 
underlying user data mechanism for use by the interface code.
As in the C~version, \UseVerb{tzero} is the initial value of the independent 
variable, and \UseVerb{u} is the vector of initial values.
\Cref{fig:cvodeinit:ocaml} uses the OCaml local open 
syntax~\cite[\textsection 7.7.7]{LeroyEtAl:OCamlMan:2018},
\verb"Cvode.(\textcdots)" and \verb"Spils.(\textcdots)", to access the 
functions \verb"Cvode.init", \verb"Cvode.Spils.solver", 
\verb"Cvode.Spils.spgmr", and
\verb"Cvode.Spils.prec_left", and also the constructors \verb"Cvode.BDF", 
\verb"Cvode.Newton", and \verb"Cvode.SStolerances".

Not all options are configured at session creation, that is, by the call to 
\verb"init".
Others are set via later calls; for instance, the residual tolerance value 
could be fine tuned by calling:
\begin{Verbatim}
Cvode.Spils.set_eps_lin cvode_mem e
\end{Verbatim}
In choosing between the two possibilities, we strove for a balance between 
enforcing correct library use and providing a simple and natural interface.
For instance, bundling the choice of Newton iteration with that of a linear 
solver and its preconditioner exploits the type system both to clarify how 
the library works and to avoid runtime errors.
Calls like that to \verb"set_eps_lin", on the other hand, are better made 
separately since the default values usually suffice and since it may make 
sense to change such settings between calls to the solver.

The example code treats a session with \cvode{}, but the \ida{} and 
\kinsol{} interfaces are similar.
The \cvodes{} and \idas{} solvers function differently.
One of the guiding principles behind the C versions of these solvers is
that their extra features 
be accessible simply by adding extra calls to existing programs and linking 
with a different library.
We respect this choice in the design of the OCaml interface.
For example, additional ``quadrature'' equations are added to the session 
created in \cref{fig:cvodeinit:ocaml} by calling:
\begin{Verbatim}
Cvodes.Quadrature.init cvode_mem fq yq
\end{Verbatim}
which specifies a function \verb"fq" to calculate the derivatives of the 
additional equations, and also their initial values in \verb"yq".
While it would have been possible to signal such enhancements in the session 
type, we decided that this would complicate rather than clarify library use, 
especially since several enhancements---namely quadratures, forward 
sensitivity, forward sensitivity with quadratures, and adjoint 
sensitivity---and their combinations are possible.
We must thus sometimes revert to runtime checks to detect whether features 
are used without having been initialized.
These checks are not always performed by the underlying library and misuse 
of certain features can give rise to segmentation faults.
The choice of whether to link Sundials/ML with the basic solver 
implementations or the enhanced ones is made during installation.

To calculate sensitivities using the adjoint method, a library user must 
first ``enhance'' a solver session using \verb"Cvodes.Adjoint.init", then 
calculate a solution by taking steps in the forward direction, before 
attaching ``backward'' sessions, and then taking steps in the opposite 
direction.
Such backward sessions are identified in Sundials by integers and the 
functions dedicated to manipulating them take as arguments a forward session 
and an integer.
But other more generic functions are shared between normal and backward 
sessions and it is necessary to first acquire a backward session pointer 
(using \verb"CVodeGetAdjCVodeBmem") before calling them.
Our interface hides these details by introducing a 
\verb"Cvodes.Adjoint.bsession" type which is realized by wrapping a standard 
session in a constructor (to avoid errors in the interface code) and storing 
the parent session and integer code along with other backward-specific 
fields as explained in \cref{sec:sessions}.

The details of the interfaces to nvectors and linear solvers are deferred 
to~\cref{sec:vectors,sec:linsolv} since the choices made and the types used 
are closely tied to the technical details of their representations in 
memory.
We mention only that, unlike Sundials, the interface enforces compatibility 
between nvectors, sessions, and linear solvers.
Such controls become more important when sensitivity enhancements are used, 
since then the vectors used are likely of differing lengths.

Sundials is a large and sophisticated library.
It is thus important to provide high-quality documentation.
For this, ocamldoc~\cite{LeroyEtAl:OCamlMan:2018} and its ability to define 
new markup tags are invaluable.
For instance, we combine the existing tags for including \LaTeX{} with the 
MathJax library\footnote{\url{http://www.mathjax.org/}} to render 
mathematical descriptions inline, and we introduce custom tags to link to 
the extensive Sundials documentation.

\subsection{Complete examples}\label{sec:example} 

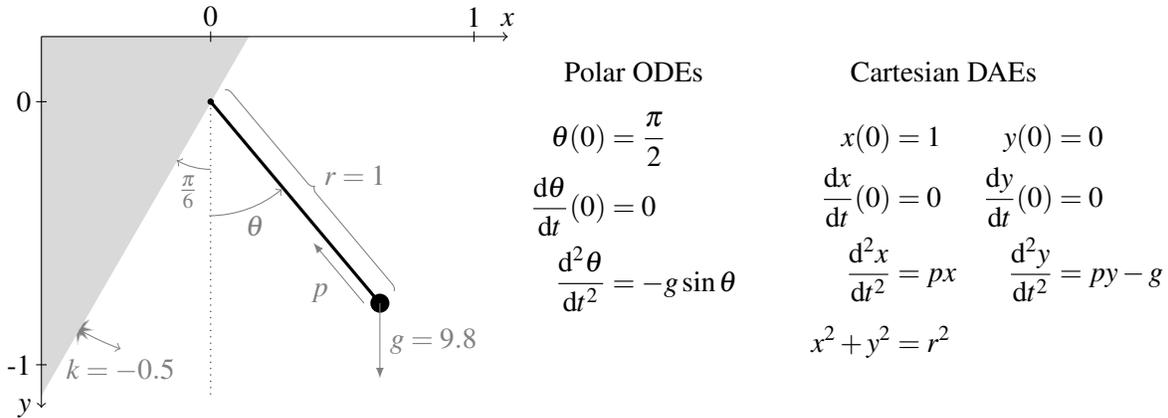
\begin{figure}
\hfil
\begin{minipage}{.4\textwidth}
  \begin{tikzpicture}[ 
      force/.style={gray,-latex},
      detail/.style={gray},
    ]
    \coordinate (attach) at (0,0);

    \path (attach) +(-120:-1) coordinate (wall top);
    \path (attach) +(-120:4.5) coordinate (wall bottom left);

    \path (attach) +(-120:3.5) coordinate (contact);
    \path (contact) node[detail,rotate=-4] {\EightStar};
    \fill[draw,gray!30]
      (wall top)
      -- (wall top-|wall bottom left)
      -- (wall bottom left)
      -- cycle;

    \draw[->]
      (wall top-|wall bottom left)
      -- +(6.2,0)
      node[above] {$x$};
    \draw[->]
      (wall top-|wall bottom left)
      -- ([yshift=-5]wall bottom left)
      node[left] {$y$};
    \draw (wall top-|attach)
      node[above] {0}
      ++(0,.07) -- +(0,-.14);
    \draw (wall top-|attach)
      ++(3.5,0) node[above] {1}
      ++(0,.07) -- +(0,-.14);

    \draw (wall bottom left|-attach)
      node[left] {0}
      ++(.07,0) -- +(-.14,0);
    \draw (wall bottom left|-attach)
      ++(0,-3.5) node[left] {-1}
      ++(.07,0) -- +(-.14,0);

    \draw[dotted] (0, 0) coordinate (attach) -- (attach|-wall bottom left);
    
    \fill[draw] (attach) circle (1pt);
    \draw[very thick,fill]
      (attach)
      -- +(-50:3.5)
      coordinate (mass)
      (mass) circle (3pt);

    \draw[force]
      ([xshift=-6]mass)
      -- node[below left] {$p$}
      ([xshift=-6]$(mass)!.3!(attach)$);
    \draw[force] (mass) -- node[right] {$g=9.8$} +(0,-1.0);

    \draw[detail,<-] (attach) +(-51:1.5cm) arc (-50:-90:1.5cm)
      node[pos=0.4,below] {$\theta$};

    \draw[detail,<-] (attach) +(-120:.9cm) arc (-120:-90:.9cm)
      node[pos=0.35,below] {$\frac{\pi}{6}$};

    \draw[detail,decorate,decoration={brace,raise=7pt}]
      (attach) -- (mass) node[midway,anchor=south west,shift={(7pt,3pt)}] 
      {$r=1$};

    \draw[detail,<->]
      (contact) arc [start angle=-120, end angle=-110, radius=3.5cm]
      node[below] {$k = -0.5$};

  \end{tikzpicture} 
\end{minipage}
\hfil
\begin{minipage}{.55\textwidth} 
  \begin{minipage}[t]{.4\textwidth}
    \centering
    Polar \acp{ODE}
    \begin{align*}
      \theta(0) &= \frac{\pi}{2} \\
      \dv{\theta}{t} (0) &= 0 \\
      \dv[2]{\theta}{t} &= -g \sin \theta
    \end{align*}
  \end{minipage}
  \hfill
  \begin{minipage}[t]{.4\textwidth}
    \centering
    Cartesian \acp{DAE}
    \begin{align*}
      x(0) &= 1 & y(0) &= 0 \\
      \dv{x}{t} (0) &= 0 & \dv{y}{t} (0) &= 0 \\
      \dv[2]{x}{t} &= px & \dv[2]{y}{t} &= py - g \\[1ex]
      x^2 + y^2 &= r^2
    \end{align*}
  \end{minipage}
\end{minipage} 
\hfil
\caption{A simple pendulum model.}\label{fig:pendulum}
\end{figure}

We now present two complete programs that use the Sundials/ML interface.
Consider the simple pendulum model shown in \cref{fig:pendulum}: a unit mass 
at the end of a steel rod is attached to an inclined wall by a friction-less 
hinge.
The rod and mass are raised parallel to the ground and released at time 
$t_0=0$.
The only forces acting on the mass are gravity and the tension from the rod.
Our aim is to plot the position of the mass as $t$ increases.
We consider two equivalent models for the dynamics: \acp{ODE} in terms of 
the angle~$\theta$ and \acp{DAE} in terms of the Cartesian coordinates~$x$ 
and~$y$.
When the mass hits the wall, its velocity is multiplied by a (negative) 
constant $k$.

\subsubsection{Polar coordinates in \cvode{}}\label{sec:example:polar} 

The \ac{ODE} model in terms of polar coordinates can be simulated with 
\cvode{}.
We start by declaring constants from the model and replacing the basic 
arithmetic operators with floating-point ones.
\begin{Verbatim}
let r, g, k = 1.0, 9.8, -0.5
and pi = 4. *. atan (1.)
and ( + ), ( - ), ( * ), ( / ) = ( +. ), ( -. ), ( *. ), ( /. )
\end{Verbatim}

The solver manipulates arrays of values, whose 0th elements 
track~$\theta$ and whose 1st elements track~$\dv{\theta}{t}$.
We declare constants for greater readability.
\begin{Verbatim}
let theta, theta' = 0, 1
\end{Verbatim}

The dynamics are specified by defining a right-hand-side function that takes 
three arguments: \verb"t", the time, \verb"y", a big array of state values, 
and \verb"yd", a big array to fill with instantaneous derivative values.
\begin{Verbatim}
let rhs t y yd =
  yd.\{theta\}  <- y.\{theta'\};
  yd.\{theta'\} <- -. g * sin y.\{theta\}
\end{Verbatim}
Apart from the imperative assignments to \verb"yd", side-effects are not 
allowed in this function, since it will be called multiple times with 
different estimates for the state values.

Interesting events are communicated to the solver through zero-crossing 
expressions.
The solver tracks the value of these expressions and signals when they 
change sign.
We define a function that takes the same inputs as the last one and that 
fills a big array \verb"r" with the values of the zero-crossing expressions.
\begin{Verbatim}
let roots t y r =
  r.\{0\} <- -. pi / 6 - y.\{theta\}
\end{Verbatim}
The single zero-crossing expression is negative until the mass collides with 
the wall.

We create a big array,\footnote{\verb"RealArray" is a helper module for 
\verb"Bigarray.Array1"s of \verb"float"s.} initialized with the initial 
state values, and ``wrap'' it as an nvector.
\begin{Verbatim}
let y = RealArray.of_list [ pi/2. ; 0. ]
let nv_y = Nvector_serial.wrap y
\end{Verbatim}
The big array, \verb"y", and the nvector, \verb"nv_y", share the same 
underlying storage.
We will access the values through the big array, but the Sundials functions 
require an nvector.

We can now instantiate the solver, using the Adams-Moulton formulas, 
functional iteration, and the default tolerance values, and staring at $t_0 
= 0$.
We pass the \verb"rhs" function defining the dynamics, the \verb"roots" 
function defining state events, and the nvector of initial states (which the 
solver uses for storage).
\begin{Verbatim}
let s = Cvode.(init Adams Functional default_tolerances
                    rhs ~roots:(1, roots) 0.0 nv_y)
\end{Verbatim}

Some solver settings are not configured through the \verb"init" routine, but 
rather by calling functions that act imperatively on the session.
Here, we set the maximum simulation time to \SI{10}{\second} and specify 
that we only care about ``rising'' zero-crossings, where a negative value 
becomes zero or positive.
\begin{Verbatim}
Cvode.set_stop_time s 10.0;
Cvode.set_all_root_directions s Sundials.RootDirs.Increasing
\end{Verbatim}
There are two main reasons for not adding these settings as optional 
arguments to \verb"init".
First and foremost, these settings may be changed during a simulation, for 
instance, to implement mode changes, so separate functions are required in 
any case.
Second, calls to \verb"init" can already become quite involved, especially 
when specifying a linear solver.
Providing optional settings in separate functions divides the interface into 
smaller pieces.
Unlike the features handled by \verb"init", the only ordering constraint 
imposed on these functions is that they be called after session 
initialization.

The simulation is advanced by repeatedly calling \verb"Cvode.solve_normal".
We define a first function to advance the simulation time~\verb"t" to a 
given value~\verb"tnext".
When a zero-crossing is signalled, it updates the array element 
for~$\theta'$, reinitializes the solver, and re-executes.
\begin{Verbatim}
let rec stepto tnext t =
  if t >= tnext then t else
  match Cvode.solve_normal s tnext nv_y with
  | (tret, Cvode.RootsFound) ->
      y.\{theta'\} <- k * y.\{theta'\};
      Cvode.reinit s tret nv_y;
      stepto tnext tret
  | (tret, _) -> tret
\end{Verbatim}

A second function calls a routine to display the current state of the system 
(and pause slightly) before advancing the simulation time by \verb"dt".
\begin{Verbatim}
let rec showloop t = if t < t_end then begin
  show (r * sin y.\{theta\}, -. r * cos y.\{theta\});
  showloop (stepto (t + dt) t)
end
\end{Verbatim}
The simulation is started by calling this function.
\begin{Verbatim}
showloop 0.0
\end{Verbatim}

Despite tail-recursive calls in \verb"stepto" and \verb"showloop", this 
program is undeniably imperative: callbacks update arrays in place, mutable 
memory is shared between arrays and nvectors, and sessions are progressively 
updated.
This is a consequence of the structure of the underlying library and works 
well in an ML language like OCaml.
We nevertheless benefit from a sophisticated type system (which infers all 
types in this example automatically), abstract data types and pattern 
matching, and exceptions.

\subsubsection{Cartesian coordinates in \ida{}}\label{sec:example:cartesian} 

\newcommand{\vx}{v_x}
\newcommand{\vy}{v_y}

The \ac{DAE} model in terms of Cartesian coordinates can be simulated with 
\ida{} once it is rearranged into the form $F(t, X, \dv{X}{t}) = 0$.
We introduce auxiliary variables~$\vx$ and~$\vy$ to represent the velocities 
and arrive at the following system with five equations and five unknowns.
\begin{align*}
  \vx - \dv{x}{t}    &= 0 & \vy - \dv{y}{t}        &= 0 &
  \dv{\vx}{t} - p x  &= 0 & \dv{\vy}{t} - p y + g  &= 0 &
  x^2 + y^2 - 1      &= 0
\end{align*}
The variable $p$ accounts for the ``pull'' of the rod.
It is the only \emph{algebraic} variable (its derivative does not appear); 
all the others are \emph{differential} variables.
The system above is of index 3, whereas an index~1 system is preferred for 
calculating the initial conditions.
The index is lowered by differentiating the algebraic constraint twice, 
giving the following equation.
\begin{align*}
  x \dv{\vx}{t} + y \dv{\vy}{t} + \vx^2 + \vy^2 &= 0
\end{align*}
Different substitutions of~$\vx$ and~$\vy$ with, respectively, $\dv{x}{t}$ 
and $\dv{y}{t}$ make for fourteen possible reformulations of the constraint 
that are equivalent in theory but that may influence the numeric solution.
Here, we choose to implement the following form.
\begin{align*}
  x \dv{\vx}{t} + y \dv{\vy}{t} + \vx \dv{x}{t} + \vy \dv{y}{t} &= 0
\end{align*}

Our second OCaml program begins with the same constant declarations as the 
first one.
This time the $X$ and $\dv{X}{t}$ arrays track five variables which we 
index,
\begin{Verbatim}
let x, y, vx, vy, p = 0, 1, 2, 3, 4
\end{Verbatim}
and there are also five residual equations to index:
\begin{Verbatim}
let vx_x, vy_y, acc_x, acc_y, constr = 0, 1, 2, 3, 4
\end{Verbatim}

The residual function itself is similar in principle to the \verb"rhs" 
function of the previous example.
It takes four arguments: \verb"t", the time, \verb"vars", a big array of 
variable values, \verb"vars'", a big array of variable derivative values, 
and \verb"res", a big array to fill with calculated residuals.
We encode fairly directly the system of equations described above.
\begin{Verbatim}
let residual t vars vars' res =
  res.\{vx_x\}   <- vars.\{vx\}   -  vars'.\{x\};
  res.\{vy_y\}   <- vars.\{vy\}   -  vars'.\{y\};
  res.\{acc_x\}  <- vars'.\{vx\}  -  vars.\{p\} * vars.\{x\};
  res.\{acc_y\}  <- vars'.\{vy\}  -  vars.\{p\} * vars.\{y\}  +  g;
  res.\{constr\} <- vars.\{x\} * vars'.\{vx\} + vars.\{y\} * vars'.\{vy\}
                  + vars.\{vx\} * vars'.\{x\} + vars.\{vy\} * vars'.\{y\}
\end{Verbatim}

The previous example applied functional iteration which does not require 
solving linear constraints.
This is not possible in \ida{}, so we will use a linear solver for which we 
choose to provide a function to calculate a Jacobian matrix for the system.
The function receives a record containing variables and their derivatives, a 
coefficient~\verb"c" that is proportional to the step size, and other values 
that we do not require.
It also receives a dense matrix~\verb"out", which we ``unwrap'' into a 
two-dimensional big array and fill with the non-zero partial derivatives of 
residuals relative to variables.
\begin{Verbatim}
let jac Ida.(\{ jac_y = vars ; jac_y' = vars'; jac_coef = c \}) out =
  let out = Matrix.Dense.unwrap out in
  out.\{x,  vx_x\}   <- -.c;             out.\{y,  vy_y\}  <- -.c;
  out.\{vx, vx_x\}   <- 1.;              out.\{vy, vy_y\}  <- 1.;
  out.\{x,  acc_x\}  <- -. vars.\{p\};     out.\{y,  acc_y\} <- -.vars.\{p\};
  out.\{vx, acc_x\}  <- c;               out.\{vy, acc_y\} <- c;
  out.\{p,  acc_x\}  <- -.vars.\{x\};      out.\{p,  acc_y\} <- -.vars.\{y\};
  out.\{x,  constr\} <- c * vars.\{vx\}  +  vars'.\{vx\};
  out.\{y,  constr\} <- c * vars.\{vy\}  +  vars'.\{vy\};
  out.\{vx, constr\} <- c * vars.\{x\}   +  vars'.\{x\};
  out.\{vy, constr\} <- c * vars.\{y\}   +  vars'.\{y\}
\end{Verbatim}

The zero-crossing function is now defined in terms of the Cartesian 
variables.

\begin{Verbatim}
let roots t vars vars' r =
  r.\{0\} <- vars.\{x\} - vars.\{y\} * (sin (-. pi / 6.) / -. cos (-. pi / 6.))
\end{Verbatim}

We create two big arrays initialized with initial values and a guess for the 
initial value of $p$, and wrap them as nvectors.
\begin{Verbatim}
let vars  = RealArray.of_list [x0; y0; 0.; 0.; 0.]
let vars' = RealArray.make 5 0.
let nv_vars, nv_vars' = Nvector.(wrap vars, wrap vars') in
\end{Verbatim}

We can now instantiate the solver session.
We create and pass a generic linear solver on 5-by-5 dense matrices 
(\verb"nv_vars" is only provided for compatibility checks) and specialize it 
for an \ida{} session with the Jacobian function defined above.
We also specify scalar relative and absolute tolerances, the residual 
function, the zero-crossing function, and initial values for the time, 
variables, and variable derivatives.
\begin{Verbatim}
let s = Ida.(init Dls.(solver ~jac (dense nv_vars (Matrix.dense 5)))
                  (SStolerances (1e-9, 1e-9))
                  residual ~roots:(1, roots) 0. nv_vars nv_vars')
\end{Verbatim}
If it were necessary to configure the linear solver or to extract statistics 
from it, we would have to declare a distinct variable for it, for example, 
\verb"let ls = Ida.Dls.dense nv_vars (Matrix.dense 5)".

Since we will be asking Sundials to calculate initial values for algebraic 
variables, that is, for $p$, we declare an nvector that classifies each 
variable as either differential or algebraic.
\begin{Verbatim}
let d, a = Ida.VarId.differential, Ida.VarId.algebraic in
let var_types = Nvector.wrap (RealArray.of_list [ d; d; d; d; a ])
\end{Verbatim}

\noindent
This information is given to the solver and algebraic variables are 
suppressed from local error tests.
\begin{Verbatim}
Ida.set_id s var_types;
Ida.set_suppress_alg s true
\end{Verbatim}

Initial values for the algebraic variables and their derivatives are then 
calculated and updated in the appropriate nvectors for a first use at time 
\verb"dt".
\begin{Verbatim}
Ida.calc_ic_ya_yd' s ~y:nv_vars ~y':nv_vars' ~varid:var_types dt
\end{Verbatim}

Then, as in the first program, we define a function to advance the 
simulation and check for zero-crossings.
If a zero-crossing occurs, the variables and derivatives are updated, the 
solver is reinitialized, and the values of algebraic variables and their 
derivatives are recalculated.
\begin{Verbatim}
  let rec stepto tnext t =
    if t >= tnext then t else
    match Ida.solve_normal s tnext nv_vars nv_vars' with
    | (tret, Ida.RootsFound) ->
        vars.\{vx\} <- k * vars.\{vx\};
        vars.\{vy\} <- k * vars.\{vy\};
        Ida.reinit s tret nv_vars nv_vars';
        Ida.calc_ic_ya_yd' s ~y:nv_vars ~y':nv_vars' ~varid:var_types (t + dt);
        stepto tnext tret
    | (tret, _) -> tret
\end{Verbatim}

The final function is almost the same as in the first program, except that 
now the state values can be passed directly to the display routine.
\begin{Verbatim}
let rec showloop t = if t < t_end then begin
  show (vars.\{x\}, vars.\{y\});
  showloop (stepto (t + dt) t)
end in
showloop 0.0
\end{Verbatim}

This second program involves more technical details than the first one, even 
if, in this case, it solves the same problem.
No new concepts are required from a programming point-of-view.

\section{Technical details}\label{sec:tech}

We now describe and justify the main typing and implementation choices made 
in the Sundials/ML interface.
For the important but standard details of writing stub functions and 
converting to and from OCaml values we refer readers to 
\Creflabel{chapter}~20 of the OCaml manual~\cite{LeroyEtAl:OCamlMan:2018}.
We focus here on the representation in OCaml of nvectors 
(\cref{sec:vectors}), sessions (\cref{sec:sessions}), and linear solvers 
(\cref{sec:linsolv}), describing some of the solutions we tried and 
rejected, and presenting the solutions.
We also describe the treatment of Jacobian matrices (\cref{sec:matrices}).

\subsection{Nvectors}\label{sec:vectors} 

Nvectors combine an array of floating-point numbers (\verb"double"s) with 
implementations of the 26 vector operations.
The OCaml big array library~\cite{LeroyEtAl:OCamlMan:2018} provides arrays 
of floating-point numbers that can be shared directly between OCaml and C.
It is the natural choice for interfacing with the payloads of nvectors.
Both Sundials' nvectors and OCaml's big arrays have a flag to indicate 
whether or not the payload should be freed when the nvector or big array is 
destroyed.

\subsubsection{An attempted solution}\label{sec:vectors:attempted} 

A first idea for an interface is to work only with big arrays on the OCaml 
side of the library, and to convert them automatically \emph{to} nvectors on 
calls into Sundials, and \emph{from} nvectors on callbacks from Sundials.
We did exactly this in early versions of our interface that only supported 
serial nvectors.\footnote{This approach is also taken in the NMAG 
library~\cite{FangohrEtAl:Nmag:2012} for interfacing with \cvode{} for both 
serial and parallel nvectors.
The Modelyze implementation~\cite{BromanSie:Modelyze:2012} provides a 
similar interface to \ida{}, but explicitly copies values to and from 
standard OCaml arrays and nvectors.}
For calls into Sundials, like \verb"Cvode.init" from the earlier example, 
the interface code creates a temporary serial nvector to pass into \cvode{} 
by calling the Sundials function
\begin{Verbatim}
N_VMake_Serial(Caml_ba_array_val(b)->dim[0], (realtype *)Caml_ba_data_val(b))
\end{Verbatim}
which creates an nvector from an existing array.
The two arguments extract the array size and the address of the underlying 
data from the big array data structure.
Operations on the resulting nvector directly manipulate the data stored
in the big array.
This nvector is destroyed by calling \emph{nvdestroy}---one of the 26 
abstract operations defined for nvectors---before returning from the 
interface code.

For callbacks into OCaml, we create a big array for each nvector argument 
\verb"v" passed in from Sundials by calling the OCaml function
\begin{Verbatim}
caml_ba_alloc(CAML_BA_FLOAT64|CAML_BA_C_LAYOUT, 1, NV_DATA_S(v), &(NV_LENGTH_S(v)))
\end{Verbatim}
From left to right, the arguments request a big array of \verb"double" 
values in row-major order indexed from 0, specify the number of dimensions, 
pass a pointer to the nvector's payload extracted using a Sundials macro, 
and give the length of the array using another Sundials macro (the last 
argument is an array with one element for each dimension).
Again, operations on the big array modify the underlying array directly.
After a callback, we set the length of the big array \verb"b" to 0,
\begin{Verbatim}
Caml_ba_array_val(b)->dim[0] = 0
\end{Verbatim}
as a precaution against the possibility, albeit unlikely, that a callback 
routine could keep a big array argument in a reference variable that another 
part of the program later accesses after the nvector memory has been freed.
The OCaml runtime will eventually garbage collect the emptied big array.
This mechanism can, however, be circumvented by creating a subarray which 
will have a distinct header, and hence dimension field, pointing to the same 
underlying memory~\cite{Bunzli:Bigarrays:2005}.

This approach has two advantages: it is simple and library users need only 
work with big arrays.
As for performance, calls into Sundials require \verb"malloc" and 
\verb"free" but they are outside critical solver loops, and, anyway, 
``wrapper'' vectors can always be cached within the session value if 
necessary.
Callbacks from Sundials do occur inside critical loops, but we can expect 
\verb"caml_ba_alloc" to allocate on the relatively fast OCaml heap and a 
\verb"malloc" is not required since we pass a data pointer.
This approach has two drawbacks: it does not generalize well to OpenMP, 
Pthreads, parallel, and custom nvectors, and a big array header is always 
allocated on the major heap which may increase the frequency and cost of 
garbage collection.
We tried to generalize this approach to handle parallel nvectors by using 
preprocessor macros, but the result was confusing for users, tedious to 
maintain, and increasingly unwieldy as we extended the interface to treat 
\cvodes{} and \idas{}.

\subsubsection{The adopted solution}\label{sec:vectors:adopted} 

The solution we finally adopted exploits features of the nvector abstract 
datatype and polymorphic typing to treat nvectors more generically and 
without code duplication.
The idea is straightforward: we pair an OCaml representation of the contents 
with a (wrapped) pointer to a C~nvector structure, and we link both to the 
same underlying data array as in the previous solution.
The difference is that we maintain both the OCaml view and the C view of the 
structure at all times.

\begin{figure}
\centering
\begin{tikzpicture}[
    node distance=1cm,
    cell/.style={draw,rectangle,thick,
                 minimum height=4.0ex,
                 minimum width=8.0em},
    wcell/.style={cell,minimum width=9.1em},
    serial/.style={thick,densely dotted},
    links/.style={thick}
  ]
    \draw[dash dot dot,gray] (2.2,-4.7) -- (2.2,1.4);
    \node[left,gray] at (2.1,1.2) {OCaml heap};
    \node[right,gray] at (2.3,1.2) {C heap};

    \node[cell] (payload) {\verb"payload"};
    \node[cell,below left=.6cm and -1cm of payload] (data) {\verb"'data"};
    \node[cell,below=-\the\pgflinewidth of data] (cnvec) {\verb"cnvec"};
    \node[cell,below=-\the\pgflinewidth of cnvec,gray]
                                        (compat) {\verb"'data -> bool"};
    \node[cell,below=-\the\pgflinewidth of data] {};
    \node[above=0 of data] {\verb"Nvector.t"};

    \node[wcell,below right=0 and 4.0 of cnvec] (content)
            {\verb"content (void *)"};
    \node[wcell,below=-\the\pgflinewidth of content] (ops) {\verb"ops"};
    \node[above=0 of content]
        {\hspace{1em}\verb"*N_Vector"\textsuperscript{+}};

    \node[wcell,dashed,below=-\the\pgflinewidth of ops]
        (backlink) {\verb"'data" (`backlink')};
    \node[below left] at (backlink.north east) {\pgfuseplotmark{square*}};

    \node[cell,serial,right=2.5cm of payload] (array)
        {\verb"double *"};
    \path (array.east) +(0.5,0) coordinate (arraylink);
    \draw[serial,->] ([xshift=1cm]payload) |- (array);
    \draw[serial,->] (content.east)
                     -| (arraylink)
                     |- (array.east)
                     ;

    \path ($(data.east)!.5!(payload.west)$) coordinate (hor);
    \draw[->,links] (cnvec) -| ([xshift=.2cm]content.north west);
    \draw[->,links] (data.east) -| (payload.south);
    \draw[->,links]
           (backlink.east)
        -- ++(1.8,0)
        -- ++(0,4.8) coordinate (ver)
        -| (payload.north)
        ;

    \node at (-2.5,-4.1) {\raisebox{.5ex}{\pgfuseplotmark{square*}}%
                                         \hspace{.7em}GC root};
\end{tikzpicture}
\caption{Interfacing nvectors (dotted lines are for serial 
nvectors).\label{fig:mem:nvec}}
\end{figure}
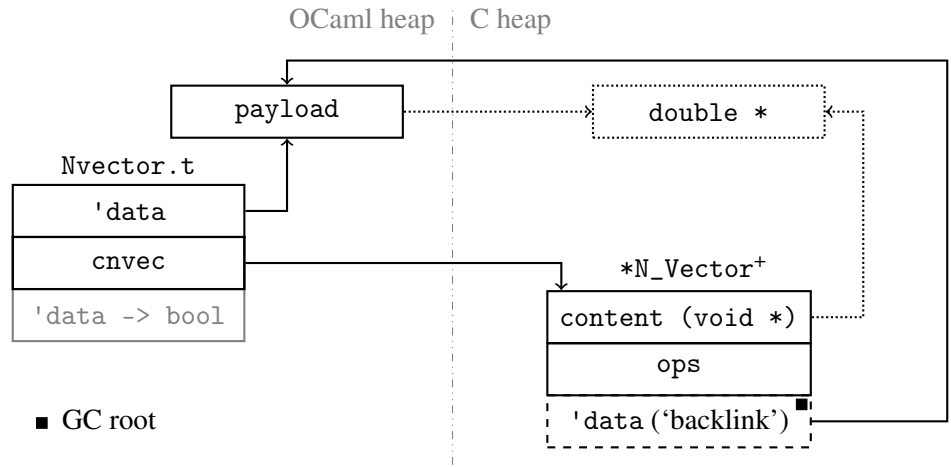

The memory layout of our nvector is shown in \cref{fig:mem:nvec}.
The OCaml type for accessing this structure is defined in the 
\verb"Nvector" module as:
\begin{Verbatim}
type ('data, 'kind) t = 'data * cnvec * ('data -> bool)
\end{Verbatim}
and used abstractly as \verb"('data, 'kind) Nvector.t" throughout the 
interface.
The \verb"'data" points to the OCaml view of the payload (labelled 
``\verb"payload"'' in \cref{fig:mem:nvec}).
For serial nvectors, \verb"'data" is instantiated as a big array of 
\verb"float"s.
The phantom type~\cite{LeijenMei:DomSpecComp:1999} argument \verb"'kind" is 
justified subsequently.
The \verb"cnvec" component is a custom block pointing to the \verb"N_Vector" 
structure allocated in the C heap.
The last component of the triple is a function that tests the compatibility 
of the nvector with another one: for serial nvectors, this means one of the 
same length, while for parallel nvectors, global sizes and \ac{MPI} 
communicators are also checked.
This component is used internally in our binding to ensure that, for 
instance, only compatible nvectors are added together.
Since the compatibility check only concerns the payload, we use a function 
from type \verb"'data" rather than type \verb"Nvector.t".
This check together with the type arguments prevents nvectors being used in 
ways that would lead to invalid memory accesses.
The two mechanisms help library users: types document how the library is 
used and dynamic checks signal problems at their point of occurrence (as an 
alternative to long debugging sessions).

With this representation, the \verb"Nvector" module can provide a generic 
and efficient function for accessing the data from the OCaml side of the 
interface:
\begin{Verbatim}
let unwrap ((payload, _, _) : ('data, 'kind) t) = payload
\end{Verbatim}
with the type \verb"('data, 'kind) Nvector.t -> 'data".
Calls from OCaml into C work similarly by obtaining a pointer to the 
underlying Sundials nvector from the \verb"cnvec" field.

Callbacks from C into OCaml require another mechanism.
Stub functions are passed \verb"N_Vector" values from which they must 
recover corresponding OCaml representations before invoking a callback.
While it would be possible to modify the nvector contents field to hold both 
an array of data values and a pointer to an OCaml value, we wanted to use 
the original nvector operations without any additional overhead.
Our solution is to allocate more memory than necessary for an 
\verb"N_Vector" so as to add a ``backlink'' field that references the OCaml 
representation.
The approach is summarised in \cref{fig:mem:nvec}.
At left are the values in the OCaml heap: an \verb"Nvector.t" and its 
\verb"'data" payload.
The former includes a pointer into the C heap to an \verb"N_Vector" 
structure extended, hence the `+', with a third field that refers back to 
the data payload on the OCaml side.
Callbacks can now easily retrieve the required value:
\begin{Verbatim}
#define NVEC_BACKLINK(nvec) (((struct cnvec *)nvec)->backlink)
\end{Verbatim}
\label{def:NVEC_BACKLINK}
The backlink field must be registered as a global root with the garbage 
collector to ensure that it is updated if the payload is moved and also that 
the payload is not destroyed inopportunely.
Pointing this global root directly at the \verb"Nvector.t" would create a 
cycle across both heaps and necessitate special treatment to avoid memory 
leaks.
We thus decided to pass payload values directly to callbacks, with the added 
advantage that callback functions have simpler types in terms of 
\verb"'data" rather than \verb"('data, 'kind) Nvector.t".
When there are no longer any references to an \verb"Nvector.t", it is 
collected and its finalizer frees the \verb"N_Vector" and associated global 
root.
This permits the payload to be collected when no other references to it 
exist.
We found that this choice works well in practice provided the standard 
vector operations are also defined for payload values---that is, directly on 
values of type \verb"'data"---since callbacks can no longer use the nvector 
operations directly.
The only drawback we experienced was in implementing custom linear solvers.
Such linear solvers take nvector arguments, which thus become payload values 
within OCaml, but they also work with callbacks into Sundials that take 
nvector arguments.
The OCaml code must thus `rewrap' payload values in order to use the 
callbacks.


The ``backlink'' field is configured when an \verb"N_Vector" is created by a 
function in the OCaml \verb"Nvector" module.
The operations of serial and other \verb"N_Vector"s are unchanged but for 
\verb"nvclone", \verb"nvdestroy", and \verb"nvcloneempty" which are replaced 
by custom versions.
The replacement \verb"nvclone" allocates and sets up the backlink field and 
associated payload.
For serial nvectors, it aliases the contents field to the data field of the 
big array payload which is itself allocated in the C heap, as shown in 
dotted lines in \cref{fig:mem:nvec}, and registers the backlink as a global 
root.
The replacement \verb"nvdestroy" removes the global root and frees memory 
allocated directly in the C heap but leaves the payload array to be garbage 
collected when it is no longer accessible.
There are thus two classes of nvectors: those created on the OCaml side with 
a lifetime linked to the associated \verb"Nvector.t" and determined by the 
garbage collector, and those cloned within Sundials, for which there is no 
\verb"Nvector.t" and whose lifetime ends when Sundials explicitly destroys 
them, though the payload may persist.

Overriding the clone and destroy operations is more complicated than the 
first attempted solution, and does not work with nvectors created outside 
the OCaml interface (which would not have the extra backlink field).
This means, in particular, that we forgo the possibility of code mixing 
Fortran, C, and OCaml, but otherwise this approach is efficient and 
generalizes well.

Within the OCaml interface, the serial, OpenMP, and Pthreads nvectors carry 
a big array payload, but at the C~level each is represented by a different 
type of struct: those for the last two, for example, track the number of 
threads in use.
OpenMP and Pthreads nvectors can be used anywhere that Serial nvectors 
can---since they are all manipulated through a common set of 
operations---except when the underlying representation is important, as in 
direct accesses to nvector data or through functions like 
\verb"Nvector_pthreads.num_threads".
We enforce these rules using the \verb"'kind" type variable introduced above 
and polymorphic variants~\cite{Garrigue:PolyVariants:1998}\cite[\textsection 
4.2]{LeroyEtAl:OCamlMan:2018}.
We use three polymorphic variant constructors, marked with backticks, and 
declare three type aliases as (closed) sets of these 
constructors:\label{page:serialkinds}
\begin{Verbatim}
type Nvector_serial.kind =   [`Serial]
type Nvector_pthreads.kind = [`Pthreads | Nvector_serial.kind]
type Nvector_openmp.kind =   [`OpenMP   | Nvector_serial.kind]
\end{Verbatim}
Here we abuse the syntax of OCaml slightly: in the real implementation, each 
kind is declared in the indicated module.
The first line declares \verb"Nvector_serial.kind" as a type whose only 
(variant) constructor is \verb"`Serial".
The second line declares \verb"Nvector_pthreads.kind" as a type whose only 
constructors are \verb"`Pthreads" and \verb"`Serial", and likewise for the 
third line.
In fact, the constructors are never used as values, since the \verb"'kind" 
argument is a `phantom'~\cite{LeijenMei:DomSpecComp:1999}: it only ever 
occurs on the left-hand side of type definitions.
They serve only to express typing constraints.
Functions that accept any kind of nvector are polymorphic in \verb"'kind", 
and those that only accept a specific kind of nvector are constrained with a 
specific \verb"kind", like one of the three listed above or others 
introduced specifically for parallel or custom nvectors.
Functions that accept any of the three kinds listed above but no others, 
since they exploit the underlying serial data representation, take a 
polymorphic nvector whose \verb"'kind" is constrained by
\begin{Verbatim}
constraint 'kind = [>Nvector_serial.kind]
\end{Verbatim}
Such an argument can be instantiated with any type that includes at least 
the \verb"`Serial" constructor.
The fact that \verb"Nvector.t" is opaque means that it can only be one of 
the nvector types \verb"Nvector_serial.t", \verb"Nvector_pthreads.t", or 
\verb"Nvector_openmp.t".
An example is given in \cref{sec:linsolv}.

For parallel nvectors, the payload is the triple:
\begin{Verbatim}
(float, float64_elt, c_layout) Bigarray.Array1.t * int * Mpi.communicator
\end{Verbatim}
where the first element is a big array of local \verb"float"s, the second 
gives the global number of elements, and the third specifies the \ac{MPI} 
processes that communicate together.\footnote{We use the OCaml MPI binding: 
\url{https://github.com/xavierleroy/ocamlmpi/}.}
We instantiate the \verb"'data" type argument of \verb"Nvector.t" with 
this triple and provide creation and clone functions that create aliasing 
for the big array and duplicate the other two elements between the OCaml and 
C representations.
A specific kind is declared for parallel nvectors.

Fur custom nvectors, we define a record type containing a field for each 
nvector operation, and a compatibility check (\verb"n_vcheck"), over an 
arbitrary payload type~\verb"'d":
\begin{Verbatim}
type 'd nvector_ops = \{
  n_vcheck     : 'd -> 'd ->  bool;
  n_vclone     : 'd -> 'd;
  n_vlinearsum : float -> 'd -> float -> 'd -> 'd -> unit;
  n_vmaxnorm   : 'd -> float;
  \vdots
\}
\end{Verbatim}
Such a record can then be used to create a wrapper function that turns 
payload values of type~\verb"'d" into custom nvectors, by calling:
\begin{Verbatim}
val make_wrap : 'd nvector_ops -> 'd -> ('d, Nvector_custom.kind) Nvector.t
\end{Verbatim}
The resulting \verb"Nvector.t" carries the type of payload manipulated by 
the given operations and a kind, \verb"Nvector_custom.kind", specific to 
custom nvectors.
The kind permits distinguishing, for instance, between a custom nvector 
whose payload is a big array of \verb"float"s and a standard serial nvector.
While there is little difference from within OCaml---both have the same type 
of payload---the differences in the underlying representations are important 
from the C~side.
In the associated \verb"N_Vector" data structure, we point the \verb"ops" 
fields at generic stub code that calls back into OCaml and store the 
closures defining the operations in the \verb"content" field.
This field is registered as a global root with the garbage collector.
The payload is stored using the backlink technique described earlier and 
depicted in \cref{fig:mem:nvec}.
It is possible to create an OCaml-C reference loop by referring back to an 
\verb"Nvector.t" value from within a custom payload or the set of custom 
operations, and thus to inhibit correct garbage collection.
This is unlikely to happen in normal use and is difficult to detect, so we 
simply rule it a misuse of the library.

\subsection{Sessions}\label{sec:sessions} 

OCaml session values must track an underlying C session pointer and also 
maintain references to callback closures and some other bookkeeping details.
We exploit the user data feature of Sundials to implement callbacks.
The main technical challenges are to avoid inter-heap loops and to smoothly 
accommodate sensitivity analysis features.

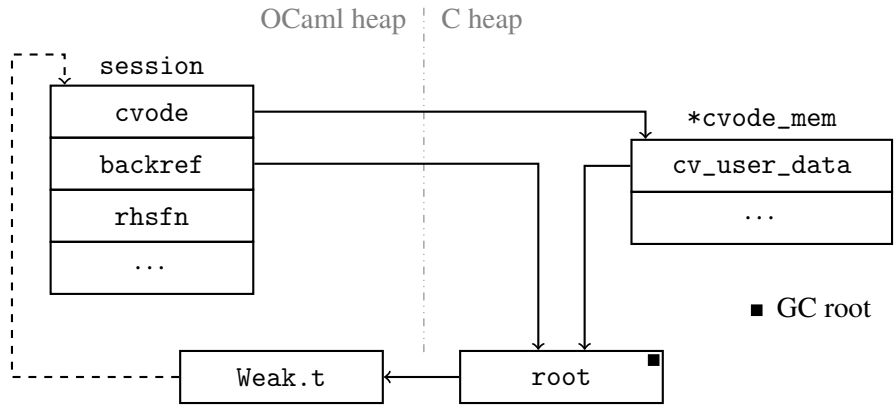
\begin{figure}
\centering
\begin{tikzpicture}[
    node distance=1cm,
    cell/.style={draw,rectangle,thick,
                 minimum height=4.0ex,
                 minimum width=7em},
    wide cell/.style={cell,minimum width=9em},
    serial/.style={thick,densely dotted},
  ]
    \draw[dash dot dot,gray] (3.615,-3.2) -- (3.615,1.4);
    \node[left,gray]  at (3.515,1.2) {OCaml heap};
    \node[right,gray] at (3.715,1.2) {C heap};

    \node[cell] (cvode) {\verb"cvode"};
    \node[cell,below=-\the\pgflinewidth of cvode] (backref) {\verb"backref"};
    \node[cell,below=-\the\pgflinewidth of backref] (rhsfn) {\verb"rhsfn"};
    \node[cell,below=-\the\pgflinewidth of rhsfn] (session etc) {\ldots};
    \node[above=0 of cvode] {\verb"session"};
    \path (cvode.north west) ++(+.2,0) coordinate (cvode link)
                             ++(-.7,.4) coordinate (cvode corner);

    \node[wide cell,below right=0cm and 5.0cm of cvode] (cvode_mem)
        {\verb"cv_user_data"};
    \node[wide cell,below=-\the\pgflinewidth of cvode_mem] (cvode_mem etc)
        {\ldots};
    \node[above=0 of cvode_mem] {\verb"*cvode_mem"};

    \draw[->,thick] (cvode) -| ([xshift=.2cm]cvode_mem.north west);

    \node[cell,below right=2\the\pgflinewidth and -1cm of session etc] (weak) 
    {\verb"Weak.t"};
    \node[cell,right=1cm of weak] (root) 
    {\verb"root"};
    \node[below left] at (root.north east) {\pgfuseplotmark{square*}};
    \draw[->,thick] (cvode_mem) -| ([xshift=.8em]root.north);
    \draw[->,thick] (backref) -| ([xshift=-.8em]root.north);
    \draw[->,thick] (root.west) -- (weak.east);

    \draw[->,thick,dashed] (weak) -| (cvode corner) -| (cvode link);

    \node at (8.8,-2.6) {\raisebox{.5ex}{\pgfuseplotmark{square*}}%
                                         \hspace{.7em}GC root};
\end{tikzpicture}
\caption{Interfacing (\cvode) sessions.\label{fig:mem:session}}
\end{figure}

The solution we implemented is sketched in \cref{fig:mem:session} and 
described below for \cvode{} and \cvodes{}.
The treatment of \ida{}, \idas{}, \arkode{}, and \kinsol{} is essentially 
the same.
As for nvectors, OCaml session types are parametrized by \verb"'data" and 
\verb"'kind" type variables.
They are represented internally as records:
\begin{Verbatim}
type ('data, 'kind) session = \{
  cvode           : cvode_mem;
  backref         : c_weak_ref;
  rhsfn           : float -> 'data -> 'data -> unit;
  \vdots
  mutable sensext : ('data, 'kind) sensext;
\}
\end{Verbatim}
The type variables are instantiated from the nvector passed to the 
\verb"init" function that creates session values.
The type variables ensure a coherent use of nvectors, which is essential in 
operations that involve multiple nvectors, like \emph{nvlinearsum}, since 
the code associated with one of the arguments is executed on the data of all 
of the arguments.

The \verb"cvode" field of the session record contains a pointer to the 
associated Sundials session value \verb"*cvode_mem".
The \verb"cv_user_data" field of \verb"*cvode_mem" is made to point at a 
\verb"malloc"ed \verb"value" that is registered with the garbage collector 
as a global root.
This root value cannot be stored directly in \verb"*cvode_mem" because 
Sundials only provides indirect access to the \verb"cv_user_data" field 
through the functions \verb"CvodeSetUserData" and \verb"CvodeGetUserData".
We would have had to violate the interface to acquire the address of the 
field in order to register it as a global root.

The root value must refer back to the \verb"session" value since it is used 
in C~level callback functions to obtain the appropriate OCaml closure.
The \verb"rhsfn" field shown in the record above is, for instance, the 
closure for the \cvode{} `right-hand side function', the \verb"f" of 
\cref{sec:overview}.
The root value stores a weak reference that is updated by the garbage 
collector if \verb"session" is moved but which does not prevent the 
destruction of \verb"session".
This breaks the cycle which exists across the OCaml and C~heaps: storing a 
direct reference in the root value would prevent garbage collection of the 
\verb"session" but the root value itself cannot be removed unless the 
\verb"session" is first finalized.
The \verb"backref" field is used only by the finalizer of \verb"session" to 
unregister the global root and to free the associated memory.

\begin{figure}
\begin{center} 
\begin{NumberedVerbatim}
static int rhsfn(realtype t, N_Vector y, N_Vector ydot, void *user_data)
\{
  CAMLparam0();\label{cbackstub:param}
  CAMLlocal2(session, r);
  CAMLlocalN(args, 3);\label{cbackstub:localn}

  WEAK_DEREF (session, *(value*)user_data);\label{cbackstub:weakderef}

  args[0] = caml_copy_double(t);\label{cbackstub:copydouble}
  args[1] = NVEC_BACKLINK(y);\label{cbackstub:backlinky}
  args[2] = NVEC_BACKLINK(ydot);\label{cbackstub:backlinkydot}

  r = caml_callbackN_exn(Field(session, RECORD_CVODE_SESSION_RHSFN), 3, args);\label{cbackstub:rhsfn}

  CAMLreturnT(int, CHECK_EXCEPTION (session, r, RECOVERABLE));\label{cbackstub:return}
\}
\end{NumberedVerbatim}
\end{center} 
\caption{Typical Sundials/ML callback stub.}\label{fig:cbackstub}
\end{figure}

\Cref{fig:cbackstub} shows the callback stub for the \verb"session.rhsfn" 
closure described above.
The C~function \verb"rhsfn()" is registered as the right-hand side function 
for every \cvode{} session created by the interface.
Sundials calls it with the value of the independent variable, \verb"t", an 
nvector containing the current value of the dependent variable, \verb"y", an 
nvector for storing the calculated derivative, \verb"ydot", and the 
session-specific pointer registered in \verb"cv_user_data".
Lines~\ref{cbackstub:param} to~\ref{cbackstub:localn} contain standard 
boilerplate for an OCaml stub function.
Line~\ref{cbackstub:weakderef} follows the references sketched in 
\cref{fig:mem:session} to retrieve a \verb"session" record: the 
\verb"WEAK_DEREF" macro contains a call to \verb"caml_weak_get".
The weak reference is guaranteed to point to a value since Sundials cannot 
be invoked from OCaml without passing the session value used in the 
callback.
Line~\ref{cbackstub:copydouble} copies the floating-point argument into the 
OCaml heap.
Lines~\ref{cbackstub:backlinky} and~\ref{cbackstub:backlinkydot} recover the 
nvector payloads using the macro described in \cref{sec:vectors}.
Line~\ref{cbackstub:rhsfn} retrieves and invokes the \verb"rhsfn" closure 
from the session object.
Finally, at line~\ref{cbackstub:return}, the return value is determined by 
checking whether or not the callback raised an exception, and if so, whether 
it was the distinguished \verb"RecoverableFailure" that signals to Sundials 
that recovery is possible.

\begin{figure}
\centering
\begin{tikzpicture}[
    node distance=1cm,
    cell/.style={draw,rectangle,thick,
                 minimum height=4.0ex,
                 minimum width=7em},
    wide cell/.style={cell,minimum width=9em},
    serial/.style={thick,densely dotted},
  ]
    \draw[dash dot dot,gray] ( 3.3,-3.0) -- (3.3, 1.3);
    \draw[dash dot dot,gray] (-1.6,-1.625) -- (0.9,-1.625);
    \node[left,gray]  at (3.2, 1.1) {OCaml heap};
    \node[right,gray] at (3.4, 1.1) {C heap};
    \node[left,gray]  at (3.2,-1.6) {OCaml stack};

    \node[cell] (cvode) {\verb"cvode"};
    \node[cell,below=-\the\pgflinewidth of cvode] (session etc) {\ldots};
    \node[above=0 of cvode] {\verb"session"};
    \path (cvode.north west) ++(+.2,0) coordinate (cvode link)
                             ++(-.7,.4) coordinate (cvode corner);

    \node[wide cell,below right=0cm and 3.4cm of cvode] (cvode_mem)
        {\verb"cv_user_data"};
    \node[wide cell,below=-\the\pgflinewidth of cvode_mem] (cvode_mem etc)
        {\ldots};
    \node[above=0 of cvode_mem] {\verb"*cvode_mem"};

    \draw[->,thick] (cvode) -| ([xshift=.2cm]cvode_mem.north west);

    \node[cell,below left=1.1cm and 2.5cm of cvode_mem] (stack)
            {\verb"argument"};
    \draw[->,thick] (cvode_mem) -| ($(cvode_mem)!.35!(stack)$)
                                |- (stack.east);

    \draw[->,thick] (stack) -| (cvode corner) -| (cvode link);
\end{tikzpicture}
\caption{Alternative session interface (not 
adopted).\label{fig:mem:altsession}}
\end{figure}
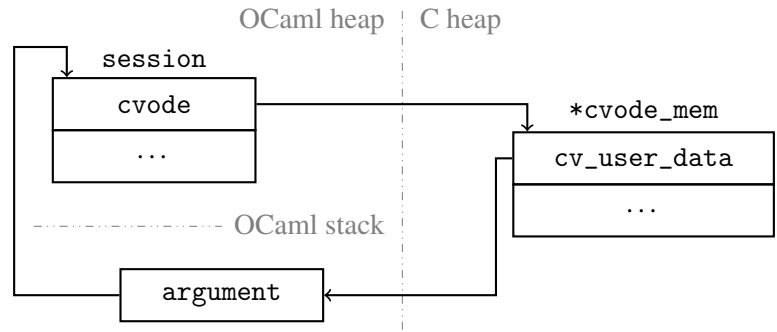


\subsubsection{An alternative session interface.}\label{sec:sessions:alt} 

Another approach for linking the C \verb"*cvode_mem" to an OCaml 
\verb"session" value is outlined in \cref{fig:mem:altsession}.
Since for a callback to occur, control must already have passed into the 
Sundials library through the interface, there will be a reference to the 
\verb"session" value on the OCaml stack.
It is thus possible to pass the reference to \verb"CvodeSetUserData" before 
calling into Sundials.
The reference will be updated by the garbage collector as necessary, but not 
moved itself during the call.
This approach is appealing as it requires neither global roots nor weak 
references.
It also requires fewer interfacing instructions in performance critical 
callback loops due to fewer indirections and because there is no need to 
call \verb"caml_weak_get".
Although this implementation is uncomplicated for functions like 
\verb"rhsfn" that are only called during solving, it is more invasive for 
the error-handling functions which can, in principle, be triggered from 
nearly every call;
either updates must be inserted everywhere or care must be taken to avoid an 
incorrect memory access.
When using an adjoint sensitivity solver, the user data references of all 
backward sessions must be updated before solving, but the error-handling 
functions do not require special treatment since they are inherited from the 
parent session.
We chose the approach based on weak references to avoid having to think 
through all such cases and also because our testing did not reveal 
significant differences in running time between the two approaches.

\subsubsection{Quadrature and Sensitivity features.}\label{sec:sessions:sens} 

Although the \cvodes{} solver conceptually extends the \cvode{} solver, it 
is implemented in a distinct code base (and similarly for \idas{} and 
\ida{}).
For the OCaml library, we wanted to maintain the idea of an extended 
interface without completely duplicating the implementation.
The library thus provides two modules, \verb"Cvode" and \verb"Cvodes", 
that share the \verb"session" type.
As both modules need to access the internals of \verb"session" values, we 
declare this type, and all the types on which it depends, inside a third 
module \verb"Cvode_impl" that the other two include.
To ensure the opacity of session types in external code, we simply avoid 
installing the associated \filename{cvode\_impl.cmi} compiled OCaml 
interface file.
The mutable \verb"sensext" field of the \verb"session" record tracks the 
extra information needed for the sensitivity features.
It has the type:
\begin{Verbatim}
type ('data, 'kind) sensext =
    NoSensExt
  | FwdSensExt of ('data, 'kind) fsensext
  | BwdSensExt of ('data, 'kind) bsensext
\end{Verbatim}
The \verb"NoSensExt" value is used for basic sessions without sensitivity 
analysis.
The \verb"FwdSensExt" value is used to augment a session with calculations 
of quadratures, forward sensitivities, and adjoint sensitivities.
It contains additional callback closures and also a list of associated 
backward session values to prevent their garbage collection while they may 
still be required by C-side callback stubs which only hold weak references.
The \verb"BwdSensExt" value is used in backward sessions created for adjoint 
sensitivity analysis.
It contains callback closures and a link to the parent session and an 
integer identifier sometimes required by Sundials functions.
Reusing the basic session interface for backward sessions mirrors the 
approach taken in Sundials and simplifies the underlying implementation at 
the cost of some redundant fields---the normal callback closures are never 
used,---and an indirection to access the \verb"bsensext" fields.

The only other complication in interfacing to \cvodes{} (and \idas{}) is 
that they often work with arrays of nvectors.
For calls into Sundials, given OCaml arrays of OCaml nvectors, we allocate 
short-lived C arrays in the interface code and extract the corresponding C 
nvector from each element.
For callbacks from Sundials, we maintain cached OCaml \verb"array"s in the 
\verb"sensext" values and populate them with OCaml payloads extracted from 
the C nvectors.

\subsection{Matrices}\label{sec:matrices} 

\begin{figure}
\centering
\begin{tikzpicture}[
    node distance=1cm,
    cell/.style={draw,rectangle,thick,
                 minimum height=4.0ex,
                 minimum width=8.0em},
    wcell/.style={cell,minimum width=9.1em},
    custom/.style={thick,densely dotted},
    links/.style={thick}
  ]
    \draw[dash dot dot,gray] (2.2,-6.8) -- (2.2,1.4);
    \node[left,gray] at (2.1,1.2) {OCaml heap};
    \node[right,gray] at (2.3,1.2) {C heap};

    \node[cell] (bigarray) {\verb"bigarray(s)"};

    \node[cell,below left=.6cm and -1cm of bigarray] (cdata) {\verb"'data"};
    \node[cell,below=-\the\pgflinewidth of cdata] (cmat) {\verb"'cmat"};
    \node[above=0 of cdata] (mcontent) {\verb"matrix_content"};

    \node[cell,below left=2.0cm and -1cm of mcontent] (mdata) {\verb"'m"};
    \node[cell,below=-\the\pgflinewidth of mdata] (mptr) {\verb"mptr"};
    \node[cell,below=-\the\pgflinewidth of mptr] (mdots) {$\cdots$};
    \node[above=0 of mdata] (matrix) {\verb"Matrix.t"};

    \node[wcell,below right=0 and 4.0 of cmat] (content)
                                                    {\verb"data (void *)"};
    \node[wcell,below=-\the\pgflinewidth of content] (dots) {$\cdots$};
    \node[above=0 of content] {\hspace{1.4em}\verb"*SUNMatrixContent"};

    \node[wcell,below right=.2 and 5.5 of mptr] (sunmat)
                                                  {\verb"content (void *)"};
    \node[wcell,below=-\the\pgflinewidth of sunmat] (cops) {\verb"ops"};
    \node[above=0 of sunmat]
        {\hspace{1em}\verb"*SUNMatrix"\textsuperscript{+}};

    \node[wcell,dashed,below=-\the\pgflinewidth of cops]
                                    (backlink) {\verb"'data" (`backlink')};
    \node[below left] at (backlink.north east) {\pgfuseplotmark{square*}};

    \node[cell,below right=0 and 3.0cm of bigarray] (array)
                                                  {\verb"double/long *"};
    \draw[links,->] ([xshift=1cm]bigarray) -| ([xshift=.2cm]array.north west);
    \draw[links,->] (content.east) -| ([xshift=1.3cm]array.south);

    \node[cell,custom,left=1.2cm of sunmat] (ops) {\verb"matrix_ops"};

    \path ($(cdata.east)!.5!(bigarray.west)$) coordinate (hor);
    \draw[->,links] (cmat) -| ([xshift=.2cm]content.north west);
    \draw[->,links] (cdata.east) -| (bigarray.south);
    \draw[->,links]
           (backlink.east)
        -- ++(1.8,0)
        -- ++(0,7.0) coordinate (ver)
        -| (mcontent.north)
        ;

    \draw[->,links] (mptr) -| ([xshift=.2cm]sunmat.north west);
    \draw[->,links] (mdata.east) -| (cmat.south);
    \draw[->,links] (mdata.east) -| (cmat.south);
    \draw[->,links] (sunmat.east) -| ([xshift=1.5cm]dots.south);

    \draw[custom,->] (sunmat) -- (ops);

    \node at (-4.0,-6.2) {\raisebox{.5ex}{\pgfuseplotmark{square*}}%
                                         \hspace{.7em}GC root};
\end{tikzpicture}
\caption{Interfacing matrices (dotted elements are for custom matrices 
only).\label{fig:mem:mat}}
\end{figure}
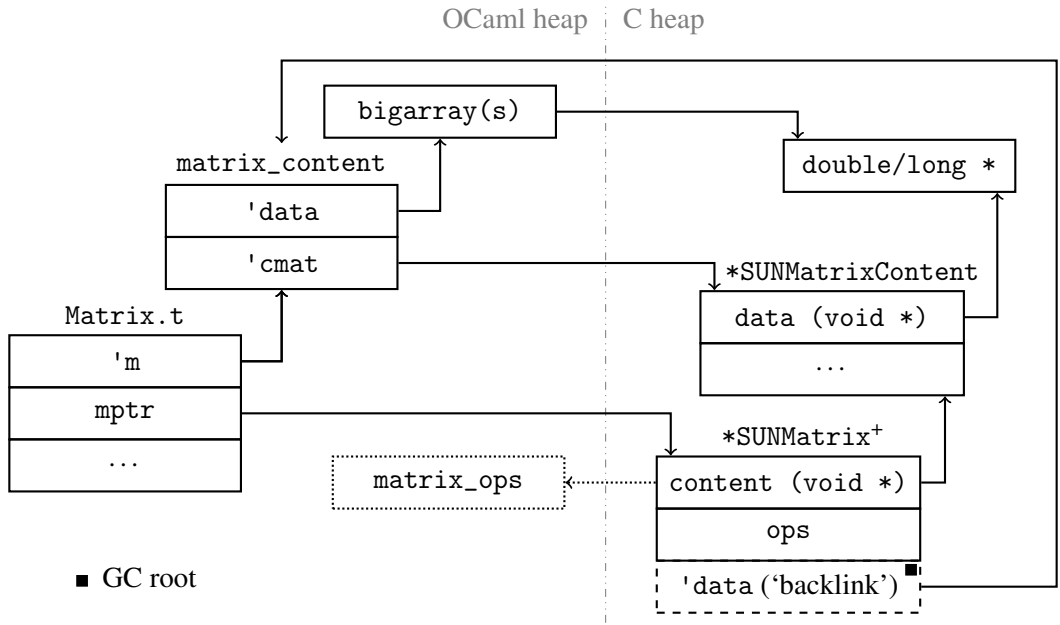

The interface for matrices must support callbacks from C into OCaml and 
allow direct access to the underlying data through big arrays.
We adapt the solution adopted for nvectors, albeit with an extra level of 
`mirroring'.
An outline of our approach is shown in \cref{fig:mem:mat}.

In C, a \verb"SUNMatrix" record pairs a pointer to content with function 
pointers to matrix operations.
The content of each matrix type is represented by a specific record type 
(\verb"SUNMatrixContent_Dense", \verb"SUNMatrixContent_Band", or 
\verb"SUNMatrixContent_Sparse"), which is not exposed to users.
The content records contain the fields described in 
\cref{sec:coverview:matrices} and pointers to underlying data arrays.

In OCaml, we implement the following types for matrices.
\begin{Verbatim}
type cmat
type ('mk, 'm, 'data, 'kind) t = \{
  payload : 'm;
  rawptr  : cmat;
  ...
\}
\end{Verbatim}
The second type is used abstractly as
\verb"('k, 'm, 'data, 'kind) Matrix.t".
The \verb"cmat" component is a custom block pointing to the \verb"SUNMatrix" 
structure allocated in the C heap; when garbage collected, its finalizer 
destroys the structure.
The \verb"payload" field of type \verb"'m" refers to the OCaml 
representation of the matrix content.
The \verb"'mk" phantom type signifies whether the underlying representation 
is the standard one provided by Sundials, in which the C-side content points 
to a \verb"SUNMatrixContent_*" structure, or a special form for custom 
matrices, in which the C-side content field is a global root referring to a 
set of OCaml closures for matrix operations.
The other two type arguments, \verb"'data" and \verb"'kind", are used to 
express restrictions on the matrix-times-vector operation.
For instance, the built-in matrix types may only be multiplied by serial, 
OpenMP, and Pthreads nvectors.

In callbacks from C into OCaml, stub functions are passed \verb"SUNMatrix" 
values from which they must recover a corresponding OCaml representation 
before invoking OCaml code.
We reuse the mechanism described in \cref{sec:vectors:adopted} for nvectors 
by adding a backlink field on the C side.
As before, this requires overriding the clone operation to recreate the 
backlink and OCaml-side structures for new matrices, and the destroy 
operation to unregister the global root.
We again prefer to avoid cross-heap cycles by not referring directly to the 
\verb"Matrix.t" wrapper.
Referring back to a big array would work well enough for dense matrices, but 
banded matrices also require tracking size and numbers of diagonals, and 
sparse matrices require the \emph{indexptrs} and \emph{indexvals} arrays 
described in \cref{sec:coverview:matrices}.

Our solution is to introduce an intermediate structure:
\begin{Verbatim}
type ('data, 'cmat) matrix_content = \{
  mutable payload : 'data;
  rawptr          : 'cmat;
\}
\end{Verbatim}
which is not exposed directly by the interface, but which is rather 
instantiated across submodules for dense, banded, sparse, and custom 
matrices.
For instance the \verb"Matrix.Dense.t" type is implemented by the following 
definitions.
\begin{Verbatim}
type data = (float, Bigarray.float64_elt, Bigarray.c_layout) Bigarray.Array2.t
type cmat
type t = (data, cmat) matrix_content
\end{Verbatim}
The \verb"cmat" represents a custom block containing a 
\verb"SUNMatrixContent_Dense" pointer to the content linked (or not) from a 
\verb"SUNMatrix" structure and the \verb"payload" field refers to a big 
array that wraps the data underlying the content structure.
Similar instantiations are used for the \verb"Matrix.Band.t" and
\verb"'s Matrix.Sparse.t" types, the latter includes a phantom type argument 
that tracks the underlying format (\ac{CSC} or \ac{CSR}).


Unfortunately, the scheme described here is made more complicated by the 
fact that certain banded and sparse operations sometimes reallocate matrix 
storage---for instance, if more non-zero elements are needed to store a 
result.
The only solution we found to this problem was to override those operations 
by duplicating the original source code and adjusting them to create new big 
arrays and link their payloads back into the C-side structures.

In a custom matrix, the matrix operations are simply overloaded by stubs 
that retrieve an OCaml closure via the \verb"SUNMatrix" content field and 
invoke it with content retrieved through the backlink.

\subsection{Linear solvers}\label{sec:linsolv} 

Sundials provides several different linear solvers and various options for 
configuring them.
One of our design goals was to clarify and ensure valid configurations using 
the OCaml module and type systems.
Our interface allows both custom and alternate linear solvers written in 
OCaml and cleanly accommodates parallel preconditioners without introducing 
a mandatory dependency on \ac{MPI}.

\subsubsection{Generic linear solvers} 

A generic linear solver essentially tries to find a vector~$x$ to satisfy an 
equation~$Ax = b$, where $A$ is a matrix and~$b$~is a vector.
Sundials provides a single type for generic linear solvers that encompasses 
instances from two families, \ac{DLS} and \ac{SPILS}.
The two families, however, are essentially implemented using different 
operations and attached to sessions by different functions with different 
supplementary arguments, for instance, \cvode{} provides 
\verb"CVDlsSetLinearSolver" and \verb"CVSpilsSetLinearSolver".
We thus found it more natural to define two distinct types, each with their 
own associated module.

Instances of the \ac{DLS} family manipulate explicit representations of the 
matrix~$A$ and the vectors~$b$ and~$x$.
The type for a generic \ac{DLS} is exposed as (slightly abusing syntax):
\begin{Verbatim}
type ('m, 'data, 'kind, 'tag) LinearSolver.Direct.linear_solver
\end{Verbatim}
where \verb"'m" captures the type of~$A$, \verb"'data" and \verb"'kind" 
constrain the nvectors for~$b$ and~$x$, and \verb"'tag" is used to restrict 
the use of operations that require KLU, SuperLUMT, or custom \ac{DLS} 
instances.
Internally, the type is realized by a record that contains a pointer to the 
C-side structure, a reference to the matrix used within Sundials for storage 
and cloning (to prevent it being prematurely garbage collected), and a 
boolean to dynamically track and prevent associations with multiple 
sessions.

\ac{DLS} implementations are provided for dense, banded, and sparse 
matrices.
The function that creates a generic linear solver over dense matrices is 
typical:
\begin{Verbatim}
  val dense : 'kind Nvector_serial.any
              -> 'kind Matrix.dense
              -> (Matrix.Dense.t, 'kind, tag) serial_linear_solver
\end{Verbatim}
The resulting generic linear solver is restricted to serial, OpenMP, or 
Pthreads nvectors:
\begin{Verbatim}
type ('m, 'kind, 'tag) serial_linear_solver
  = ('m, Nvector_serial.data, [>Nvector_serial.kind] as 'kind, 'tag) linear_solver
\end{Verbatim}
The \verb"[>Nvector_serial.kind]" constraint only allows the type variable 
\verb"'kind" to be instantiated by a type that includes the constructor 
\verb"Nvector_serial.kind", which was presented on 
\cpageref{page:serialkinds}.

A custom \ac{DLS} implementation is created by defining a record of OCaml 
functions:
\begin{Verbatim}
type ('m, 'data, 's) ops = \{
  init  : 's -> unit;
  setup : 's -> 'm -> unit;
  solve : 's -> 'm -> 'data -> 'data -> unit;
  get_work_space : ('s -> int * int) option;
\}
\end{Verbatim}
where \verb"'s" is the type of the internal state of an instance.
The following two functions are provided.
\begin{Verbatim}
val make : ('m, 'data, 's) ops -> 's -> ('mk, 'm, 'data, 'kind) Matrix.t
           -> ('m, 'data, 'kind, [`Custom of 's]) linear_solver
val unwrap : ('m, 'data, 'kind, [`Custom of 's]) linear_solver -> 's
\end{Verbatim}
The first takes a set of operations, an initial state, and a storage matrix, 
and returns a \ac{DLS} instance.
The matrix kind~\verb"'mk", indicating whether the matrix implementation is 
standard or custom, need not be propagated to the result type since 
callbacks only receive the matrix contents of type~\verb"'m".
The tag type argument indicates a custom linear solver whose internal state 
has the given type.
This tag allows for a generic \verb"unwrap" function and thereby avoids 
requiring users to maintain both a custom state---to set properties or get 
statistics---and an actual instance.

Instances of the \ac{SPILS} family rely on a function that approximates a 
matrix-vector product to model the (approximate) effect of the matrix~$A$ 
without representing it explicitly.
The success of this approach typically requires the solving a preconditioned 
system that results from scaling the matrix~$A$ and vectors~$b$ and~$x$, and 
multiplying them by problem-specific matrices on the left, right, or both 
sides.

\filbreak
The type for a generic \ac{SPILS} is exposed as:
\begin{Verbatim}
type ('data, 'kind, 'tag) LinearSolver.Iterative.linear_solver
\end{Verbatim}
where \verb"'data" and \verb"'kind" constrain the nvectors used and 
\verb"'tag" is used to restrict operations for specific iterative methods.
The internal realization of this type and the creation of custom linear 
solvers is essentially the same as for the \ac{DLS} module.
The following \ac{SPILS} instantiation function is typical.
\begin{Verbatim}
val spgmr : ?maxl:int
            -> ?max_restarts:int
            -> ?gs_type:gramschmidt_type
            -> ('data, 'kind) Nvector.t
            -> ('data, 'kind, [`Spgmr]) linear_solver
\end{Verbatim}
It takes three optional arguments to configure the linear solver and an 
nvector to specify the problem size and compatibility constraints.

\subsubsection{Associating generic linear solvers to sessions} 

Generic linear solvers are associated with sessions after having created the 
session and before simulating it.
In the OCaml interface, we incorporated these steps into the \verb"init" 
functions that return solver sessions, as shown in 
\cref{fig:cvodeinit:ocaml} and applied in \cref{sec:example:cartesian}.
This allows us to enforce required typing constraints and ensure that calls 
to the underlying library are made in the correct order.
We introduce intermediate types to represent the combination of a generic 
linear solver with its session-specific parameters and to group diagonal 
approximation, \ac{DLS}, \ac{SPILS}, and alternate modules.
For instance, in the \cvode{} solver, we declare:
\begin{Verbatim}
type ('data, 'kind) session_linear_solver
\end{Verbatim}
which is realized internally by a function over a session value and an 
nvector, and acts imperatively to configure the session.
Values of this type are provided by solver-specific submodules whose 
particularities we now summarize.
For our purposes, the important thing is not what the different modules do, 
but rather how the constraints on their use are expressed in Sundials/ML.

\paragraph{Diagonal linear solvers.}

The \cvode{} diagonal linear solver is interfaced by the submodule:
\begin{Verbatim}
module Diag : sig
  val solver : ('data, 'kind) session_linear_solver
  val get_work_space : ('data, 'kind) session -> int * int
  val get_num_rhs_evals : ('data, 'kind) session -> int
end
\end{Verbatim}
A \verb"Cvode.Diag.solver" value is passed to \verb"Cvode.init" or 
\verb"Cvode.reinit" where it is invoked and makes calls to the Sundials 
\verb"CVodeSetIterType" and \verb"CVDiag" functions that set up the diagonal 
linear solver.
The \verb"get_"$\ast$ functions retrieve statistics specific to the diagonal 
linear solver---here the memory used by the diagonal solver in terms of real 
and integer values, or the number of times that the right-hand side callback 
has been invoked.
Other linear solvers also provide \verb"set_"$\ast$ functions.
As the underlying implementations of these functions sometimes typecast 
memory under the assumption that the associated linear solver is in use, we 
implement dynamic checks that throw an exception when a session is 
configured with one linear solver and passed to a function that assumes 
another.
This constraint cannot be adequately expressed using static types since a 
session may be dynamically reinitialized with a different linear solver.


\paragraph{\acfp{DLS}.}

Interfacing \acp{DLS} requires treating nvector compatibility and the matrix 
data structures passed to callback functions.
For instance, the \verb"Cvode.Dls" submodule contains the value:
\begin{Verbatim}
val solver : ?jac:'m jac_fn ->
             ('m, 'kind, 'tag) LinearSolver.Direct.serial_linear_solver ->
             'kind serial_session_linear_solver
\end{Verbatim}
where \verb"serial_session_linear_solver" is another abbreviation for 
restricting nvectors.
\begin{Verbatim}
type 'kind serial_session_linear_solver =
  (Nvector_serial.data, [>Nvector_serial.kind] as 'kind) session_linear_solver
\end{Verbatim}
The \verb"?jac" label marks a named optional argument.
It is used to pass a function that calculates an explicit representation of 
the system Jacobian matrix.
The only relevant detail here is the use of \verb"'m" to ensure that the 
same matrix type is used by the callback function and the generic linear 
solver.
Similarly, \verb"'kind" is propagated to the result type to ensure nvector 
compatibility when a session is created.

Each solver has its own \verb"Dls" submodule into which the types and values 
of \verb"LinearSolver.Direct" are imported to maximize the effect of the 
local open syntax---for instance, as in the call to \verb"Ida.init" in the 
example of \cref{sec:example:cartesian}.

\paragraph{\acfp{SPILS}.}

A \ac{SPILS} associates an iterative method with a preconditioner.
Iterative methods are exposed as functions that take an optional Jacobian 
multiplication function and a preconditioner, for example,
\begin{Verbatim}
val solver :
  ('data, 'kind, 'tag) LinearSolver.Iterative.linear_solver
  -> ?jac_times_vec:'data jac_times_setup_fn option * 'data jac_times_vec_fn
  -> ('data, 'kind) preconditioner
  -> ('data, 'kind) session_linear_solver
\end{Verbatim}
As is clear from the type signature, it is the preconditioner that 
constrains nvector compatibility.
Internally the \verb"preconditioner" type pairs a preconditioning ``side'' 
(left, right, both, or none) with a function that configures a 
preconditioner given a session and an nvector.
Functions are provided to produce elements of this type.
For instance, the \verb"Cvode.Spils" module provides:
\begin{Verbatim}
val prec_none : ('data, 'kind) preconditioner
val prec_left  : ?setup:'data prec_setup_fn
                 -> 'data prec_solve_fn
                 -> ('data, 'kind) preconditioner
val prec_right : ?setup:'data prec_setup_fn
                 -> 'data prec_solve_fn
                 -> ('data, 'kind) preconditioner
val prec_both  : ?setup:'data prec_setup_fn
                 -> 'data prec_solve_fn
                 -> ('data, 'kind) preconditioner
\end{Verbatim}
The last three produce preconditioners from optional setup functions and 
mandatory solve functions over the nvector payload type.
These preconditioners are compatible with any type of nvector.

Banded preconditioners, on the other hand, are only compatible with serial, 
OpenMP, and Pthreads nvectors.
We group them into a submodule \verb"Cvode.Spils.Banded":
\begin{Verbatim}
val prec_left  : bandrange
                 -> (Nvector_serial.data, [> Nvector_serial.kind]) preconditioner
val prec_right : bandrange
                 -> (Nvector_serial.data, [> Nvector_serial.kind]) preconditioner
val prec_both  : bandrange
                 -> (Nvector_serial.data, [> Nvector_serial.kind]) preconditioner
\end{Verbatim}
The banded preconditioners provide their own setup and solve functions.

The \acf{BBD} preconditioner is only compatible with parallel nvectors.
Its \verb"'data" type variable is instantiated to 
\verb"Nvector_parallel.data", the payload of parallel nvectors, and its 
\verb"'kind" type variable to \verb"Nvector_parallel.kind".
The declarations are made in a separate module \verb"Cvode_bbd", which is 
simply not compiled when \ac{MPI} is not available.

Each solver has a \verb"Spils" submodule into which the types and values of 
\verb"LinearSolver.Iterative" are imported to maximize the effect of the 
local open syntax---as shown, for instance, in \cref{fig:cvodeinit:ocaml}.

\section{Evaluation}\label{sec:eval} 

An interface layer inevitably adds run-time overhead: there is extra code to 
execute at each call to, or callback from the library.
This section presents our attempt to quantify this overhead.
Since we are interested in the cost of using Sundials from programs written 
in OCaml, rather than count the number of additional instructions per call 
or callback we think it more relevant to compare the performance of programs 
written in OCaml with equivalent programs written directly in C.
We consider two programs equivalent when they 
produce identical sequences of output bytes using the same sequence of 
solver steps in the Sundials library.
Here we compare wall clock run times, which, despite the risk of 
interference from other processes and environmental factors, have the 
advantages of being relatively simple to measure and directly relevant to 
users.

Sundials is distributed with example programs (71 with serial, OpenMP, or 
Pthreads nvectors and 21 with parallel nvectors---not counting duplicates) 
that exercise a wide range of solver features in numerically interesting 
ways.
We translated them all into OCaml.\footnote{The translations aim to 
facilitate direct comparison with the original code, to ease debugging and 
maintenance. They are not necessarily paragons of good OCaml style.}
Comparing the outputs of corresponding OCaml and C versions with 
\texttt{diff} led us to correct many bugs in our interface and example 
implementations, and even to discover and report several bugs in Sundials 
itself.
We also used \verb"valgrind"~\cite{NethercoteSew:Valgrind:2007} and manual 
invocations of the garbage collector to reveal memory-handling errors in our 
code.

\SaveVerb{unsafe}"--unsafe"
\SaveVerb{buildtype}"CMAKE_BUILD_TYPE=Release"

\begin{figure*}
\begin{center}
\includegraphics[height=.71\textwidth,angle=270,clip,trim=0 0 0 0]{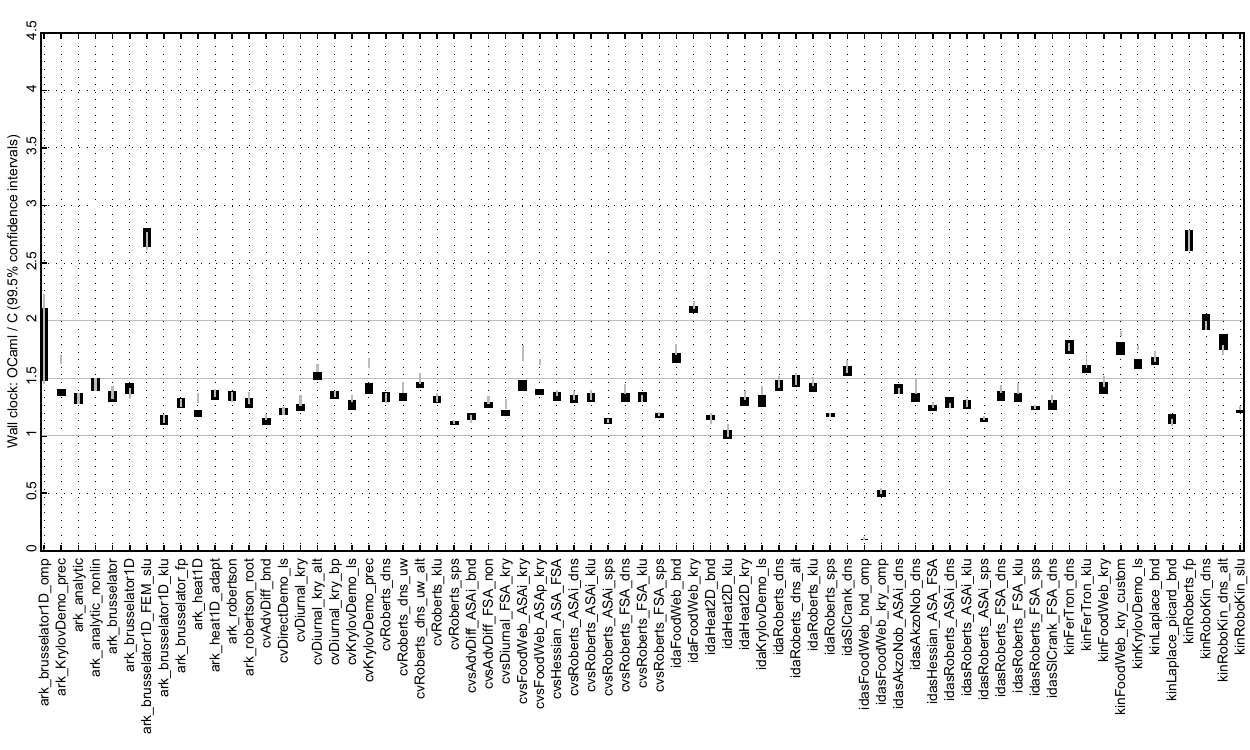}
\end{center}
\caption{Serial examples: C (gcc 6.3.0 with -O3) versus
OCaml native code (4.07.0).
The black bars show results obtained with the \protect\UseVerb{unsafe} 
option that turns off array bounds checking and other dynamic checks.
The grey lines show the results with dynamic checks.
Results were obtained under Linux 4.9.0 on an Intel i7 running at 
\SI{2.60}{\GHz} with a \SI{1}{\mega\byte} L2 cache, a \SI{6}{\mega\byte} L3 
cache, and \SI{8}{\giga\byte} of RAM.
Sundials was compiled with 
\protect\UseVerb{buildtype}.}\label{fig:benchmarks:serial}
\end{figure*}

\begin{figure*}
\begin{center}
\includegraphics[height=.71\textwidth,angle=270,clip,trim=0 0 246 
0]{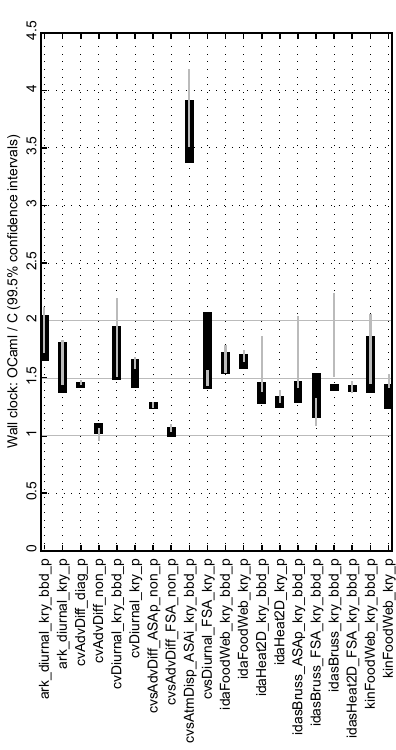}
\end{center}
\caption{Parallel examples: C (gcc 6.3.0 with -O3) versus
OCaml native code (4.07.0).
The black bars show results obtained with the \protect\UseVerb{unsafe} 
option that turns off array bounds checking and other dynamic checks.
The grey lines show the results with dynamic checks.
Results were obtained under Linux 4.9.0 on an Intel i7 running at 
\SI{2.60}{\GHz} with a \SI{1}{\mega\byte} L2 cache, a \SI{6}{\mega\byte} L3 
cache, and \SI{8}{\giga\byte} of RAM.
Sundials was compiled with 
\protect\UseVerb{buildtype}.}\label{fig:benchmarks:parallel}
\end{figure*}

After ensuring the equivalence of the example programs, we used them to 
obtain and optimize performance results.
As we explain below, most optimizations were in the example programs 
themselves, but we were able to validate and evaluate some design choices, 
most notably the alternative proposal for sessions described in 
\cref{sec:sessions}.
The bars in \cref{fig:benchmarks:serial,fig:benchmarks:parallel} show the 
ratios of the execution times of the OCaml code against the C 
code.\footnote{The Sundials/ML source code distribution includes all examples 
  along with the scripts and build targets necessary to reproduce the 
  experiments described in this paper.}
A value of 2.0 on the horizontal axis means that the OCaml version takes 
twice as long as the C version to calculate the same result.

The extent of the bars show 99.5\% confidence intervals for the OCaml/C 
ratio, calculated according to Chen~et~al.'s 
technique~\cite{ChenEtAl:IEEETrans:2015}.
Formally, if $O$ and $C$ are random variables representing the running time 
of a given test in OCaml and C, respectively, then the bars show the range 
of all $\gamma$ for which the Mann-Whitney U-test (also known as the 
Wilcoxon rank-sum test) does \emph{not} reject the null hypothesis
$
  P(\gamma C > O) = P(\gamma C < O)
$
at the 99.5\% confidence level.
Intuitively, if we handicap the C code by scaling its time by $\gamma$, then 
the observed measurements are consistent with the assumption that the 
handicapped C code beats the non-handicapped OCaml code exactly half of the 
time: the observed data may be skewed in one direction or the other, but the 
deviation from parity is smaller than random noise would give 99.5\% of the 
time if they really were equally competitive.

The results include customized examples: the 
\filename{kinFoodWeb\_kry\_custom} example uses custom nvectors with 
low-level operations implemented in OCaml on \verb"float array"s; the 
\filename{$\ast$\_alt} examples use an alternate linear solver reimplemented 
in OCaml using the underlying \verb"Dls" binding.
This involves calls from OCaml to C to OCaml to C.
Each custom example produces the same output as the corresponding original 
(their predecessors in the graph).
Two OCaml versions, \filename{idasFoodWeb\_bnd\_omp} and 
\filename{idasFoodWeb\_kry\_omp} are faster than the C versions---we explain 
why below.
The black bars in \cref{fig:benchmarks:serial,fig:benchmarks:parallel} give 
the ratios achieved when the OCaml versions are compiled without checks on 
array access, nvector compatibility, or matrix validity (since the C code 
does not perform these checks).
The grey lines show the results with dynamic checks; the overhead is 
typically negligible.
The graph suggests that the OCaml versions are rarely more than 50\% slower 
than the original ones and that they are often less than 20\% slower.
That said, care must be taken extrapolating from the results of this 
particular experiment to the performance of real applications, which will 
not be iterated \num{1000}s of times and where the costs of garbage 
collection can be minimized given sufficient memory.

The actual C run times are not given in 
\cref{fig:benchmarks:serial,fig:benchmarks:parallel}.
Most of them are less than \SI{1}{\milli\s}, nearly all of them are less 
than \SI{1}{\s}, the longest is on the order of \SI{4}{\s} 
(\filename{ark\_diurnal\_kry\_bbd\_p}).
We were not able to profile such short run times directly: the 
\filename{time} and \filename{gprof} commands simply show \SI{0}{\s}.
The figures in the graph were obtained by modifying each example (in both C 
and OCaml) to repeatedly execute its \verb"main" function.
Since the C code calls \verb"free" on all of its data, we also manually 
trigger a full major collection and heap compaction in OCaml at the end of the 
program (the ratios are smaller otherwise).
The figure compares the median ratios over 10 such executions.
The fastest examples require \num{1000}s of iterations to produce a 
measurable result, so we must vary the number of repetitions per example to 
avoid the slower examples taking too long (several hours each).
Iterating the examples so many times sometimes amplifies factors other than 
interface overhead.

For the two examples where OCaml apparently performs better than C, the 
original C~code includes OpenMP pragmas around loops in the callback 
functions.
This actually slows them down and the OCaml code does better because this 
feature is not available.

In general, we were able to make the OCaml versions of the examples up to 
four times faster by following three simple and unsurprising guidelines.
\begin{enumerate}
\item
We added explicit type annotations to all vector arguments.
For instance, rather than declare a callback with
\begin{Verbatim}
let f t y yd = ...
\end{Verbatim}
we follow the standard approach of adding type annotations,
\begin{Verbatim}
type rarray = (float, float64_elt, c_layout) Bigarray.Array1.t
let f t (y : rarray) (yd : rarray) = ...
\end{Verbatim}
so that the compiler need not generate polymorphic code and can optimize for 
the big array layout.

\item
We avoided the functions \verb"Bigarray.Array1.sub", 
\verb"Bigarray.Array2.slice_left" that allocate fresh big arrays on the 
major heap and thereby increase the frequency and cost of garbage 
collection.
They can usually be avoided by explicitly passing and manipulating array 
offsets.
We found that when part of an array is to be passed to another function, as 
for some \ac{MPI} calls, it can be faster to copy into and out of a 
temporary array.

\item
We wrote numeric expressions and loops according to 
Leroy's~\cite{Leroy:Numeric:2002} advice to avoid float `boxing'.

\end{enumerate}

As a side benefit of the performance testing, iterating each example program 
\num{1000}s of times with some control over the garbage collector revealed 
several subtle memory corruption errors in our interface.
We investigated and resolved these using manual code review and a 
C~debugger.

In summary, the results obtained, albeit against an admittedly small set of 
examples, indicate that OCaml code using the Sundials solvers should rarely 
be more than 50\% slower than equivalent code written in C, provided the 
guidelines above are followed, and it may be only about 20\% slower.
One response to the question ``Should I use OCaml to solve my numeric 
problem?'' is to rephrase it in terms of the cost of calculating results 
(``How long must I wait?'') against the cost of producing and maintaining 
programs (``How much effort will it take to write, debug, and later modify a 
program?'').
This section provides insight into the former cost.
The latter cost is difficult to quantify, but it is arguably easier to write 
and debug OCaml code thanks to automatic memory management,
strong static type checking,
bounds checking on arrays,
and higher-order functions.
This combined with the other features of OCaml---like algebraic data types, 
pattern matching, and polymorphism---make the Sundials/ML library especially 
compelling for programs that combine numeric calculation and symbolic 
manipulation.

\section{Conclusion}\label{sec:concl} 

We present a comprehensive OCaml interface to the Sundials suite of numeric 
solvers.
We outline the main features of the underlying library, demonstrate its use 
via our interface, describe the implementation in detail, and summarize 
extensive benchmarking.

Sundials is an inherently imperative library.
Data structures like vectors and matrices are reused to minimize memory 
allocation and deallocation, and modified in place to minimize copying.
It works well in an ML-style language where it is possible to mix features 
of imperative programming---like sequencing, loops, mutable data structures, 
and exceptions---with those of functional programming---like higher-order 
functions, closures, and algebraic data types.
An interesting question that we did not treat is how to build efficient 
numerical solvers in a more functional style.

It turns out that the abstract data types used to structure the Sundials 
library are also very useful for implementing a high-level interface.
By overriding elements of these data structures, namely the clone and 
destroy operations, we are able to smoothly integrate them with OCaml's 
automatic memory management.
Designers of other C libraries that intend to support high-level languages 
might also consider this approach.
For Sundials, some minor improvements are possible---for instance, adding 
\begin{inparaenum}
\item a user data field to nvectors and matrices that could be exploited for 
  `backlinks',
\item a function to return the address of the session user data field so 
  that it can be registered directly with the garbage collector, and
\item a mechanism for overriding the reallocation mechanism within banded 
  and sparse matrices to eliminate the need to reimplement certain 
  operations,
\end{inparaenum}---but the approach works well overall.

In our interface, C-side structures are mirrored by OCaml structures that 
combine low-level pointers, their associated finalize functions, and 
high-level values.
This is a standard approach that we adapted for two particular requirements, 
namely, to give direct access to low-level arrays and to treat callbacks 
efficiently.
The first requirement is addressed by combining features of the OCaml big 
array library with the ability to override the Sundials clone and destroy 
operations.
The second requirement necessitates a means to obtain the OCaml ``mirror'' 
of a given instance of a C data structure.
The backlink fields solve this problem well provided care is taken to avoid 
inter-heap cycles.
Besides the usual approach of using weak references, we demonstrate an 
alternative for when the mirrored structures are ``wrappers''.
In this case, the pointer necessary to recover an OCaml structure from C 
drops a level in the wrapper hierarchy.
This is a simple solution to the cycle problem that is also convenient for 
library users.
We note that it engenders two kinds of instance: those created in OCaml and 
those cloned in C.
For instances created in OCaml and used in C, care must be taken to maintain 
a reference from OCaml until the instance is no longer needed as the 
backlink itself does not prevent garbage collection and ensuing instance 
destruction.
For instances created in C and passed back into OCaml, there is no such 
problem provided the destruction function only unregisters the global root 
and does not actually destroy the instance.

Our work relies heavily on several features of the OCaml runtime, most 
notably flexible functions for creating and manipulating big arrays, the 
ability to register global garbage collection roots in structures on the C 
heap, features for invoking callbacks and recovering exceptions from them, 
and the ability to associate finalize functions to custom blocks.
We found phantom types combined with polymorphic variants are a very 
effective way to express constraints on the use and combination of data 
structures that arise from the underlying library.
While \acp{GADT} are used in our implementation, they are not exposed 
through the interface, probably because in this library we prefer opaque 
types to variant types since they permit a better division into submodules.

The evaluation results show that the programming convenience provided by 
OCaml and our library has a measurable runtime cost.
The overall results, however, are not as bad as may perhaps be feared.
We conjecture that, in most cases, the time gained in programming, 
debugging, and maintenance will outweigh the cumulative time lost to 
interface overhead.
Since programs that build on Sundials typically express sophisticated 
numerical models, ensuring their readability through high-level features and 
good style is an important consideration.
Ideally, programming languages and their libraries facilitate reasoning 
about models, even if only informally, and precisely communicating 
techniques, ideas, and knowledge.
We hope our library contributes to these aims.

\section*{Acknowledgements}\label{sec:ack}

We thank Kenichi Asai and Mark Shinwell for their persistence and 
thoroughness, and the anonymous reviewers for their helpful suggestions.
The work described in this paper was supported by the ITEA 3 project 11004 
MODRIO (Model driven physical systems operation).

\bibliography{paper}
\end{document}